\documentclass[a4paper,fleqn]{cas-dc}

\usepackage[numbers]{natbib}
\usepackage{amsmath}
\usepackage{amssymb}
\usepackage{booktabs}
\usepackage{multirow}
\usepackage{tabularx}
\usepackage{algorithm}
\usepackage{algorithmicx}
\usepackage{algpseudocode}
\usepackage{color}

\begin{document}
\let\WriteBookmarks\relax
\def\floatpagepagefraction{1}
\def\textpagefraction{.001}

\shorttitle{Scenario-conditioned flow matching for three-component ground-motion waveforms}
\shortauthors{Yi Ding et~al.}

\title[mode = title]{Scenario-conditioned flow matching for probabilistic generation of three-component ground-motion waveforms}

\author[1]{Yi Ding}
\author[1,2]{Jinjun Hu}
\ead{hu-jinjun@163.com}
\cormark[1]
\author[3]{Su Chen}
\author[4]{Xianwei Liu}
\author[5]{Zhongxiang Zhang}
\author[6]{Zongchao Li}
\author[3,6]{Xiaojun Li}
\author[1,2]{Lili Xie}

\affiliation[1]{organization={State Key Laboratory of Precision Blasting, Jianghan University}, city={Wuhan}, postcode={430056}, country={China}}
\affiliation[2]{organization={Key Laboratory of Earthquake Engineering and Engineering Vibration, Institute of Engineering Mechanics, China Earthquake Administration}, city={Harbin}, postcode={150080}, country={China}}
\affiliation[3]{organization={Key Laboratory of Urban Security and Disaster Engineering of the Ministry of Education, Beijing University of Technology}, city={Beijing}, postcode={100124}, country={China}}
\affiliation[4]{organization={School of Civil Engineering and Architecture, East China Jiaotong University}, city={Nanchang}, postcode={330013}, country={China}}
\affiliation[5]{organization={Key Laboratory of Roads and Railway Engineering Safety Control, Ministry of Education, Shijiazhuang Tiedao University}, city={Shijiazhuang}, postcode={050043}, country={China}}
\affiliation[6]{organization={Institute of Geophysics, China Earthquake Administration}, city={Beijing}, postcode={100081}, country={China}}

\cortext[cor1]{Corresponding author}

\begin{abstract}
Performance-based seismic risk assessment requires three-component acceleration histories compatible with specified source, path, and site conditions. Conventional ground-motion prediction equations provide scalar intensity measures, while many generative waveform models learn amplitude and waveform shape within a single high-dimensional target. We present WaveFlowGMM, a two-stage probabilistic ground-motion model that uses peak ground acceleration (PGA) as an amplitude interface between scenario conditioning and waveform generation. The amplitude stage uses physics-informed symbolic learning to estimate component-wise PGA medians and a full cross-component covariance. The waveform stage uses few-step AlphaFlow in a reconstructible wavelet-packet coefficient space to generate normalised three-component histories that are rescaled by sampled PGA. Tests on an event-level NGA-West2 holdout set show that the generated motions reproduce the main magnitude, distance, and site scaling, retain near-zero peak and spectral residuals, and preserve cross-component amplitude dependence. The framework provides candidate waveform generation components that can be further incorporated into waveform-level seismic hazard and risk analysis.
\end{abstract}

\begin{highlights}
\item A two-stage probabilistic framework separates amplitude sampling from waveform-shape generation.
\item An interpretable PGA model estimates a full cross-component covariance to preserve three-component amplitude dependence.
\item Few-step AlphaFlow samples three-component waveforms faster than the tested diffusion models. 
\end{highlights}


\maketitle

\section{Introduction}

Performance-based seismic design and seismic risk assessment require three-component acceleration histories for nonlinear time-history analysis. These records must be compatible with the source, path, and site conditions of an engineering scenario while preserving phase, duration, envelope, and component dependence. In the PEER performance-based earthquake engineering framework, engineering demand parameters and damage states are therefore evaluated from time histories, not only from scalar response-spectrum medians. Large strong-motion databases such as NGA-West2 \citep{Ancheta_2014_NGAWest2Database} provide the empirical basis for this practice, but their coverage remains uneven in the joint scenario space. When magnitude, distance, site condition, and spectral demand are prescribed together, the number of compatible observed records can be too small for direct record selection \citep{Bommer_2000_HazardconsistentEarthquakeScenarios,Luco_2007_StructurespecificScalarIntensity}.

Current engineering workflows combine scalar ground-motion models with record selection, record modification, or numerical simulation. The NGA-West2 ground-motion prediction equations \citep{Abrahamson_2014_SummaryASK14Ground,Boore_2014_NGAWest2EquationsPredicting,Campbell_2014_NGAWest2GroundMotion,Chiou_2014_UpdateChiouYoungs} provide well-calibrated medians and dispersions for PGA, PGV, and 5\%-damped PSA. Their outputs are scalar ordinates, so the phase, duration, envelope, and cross-component structure required by nonlinear time-history analysis must come from another source. Spectrum-matching methods reshape seed records in the frequency domain \citep{Scanlan_1974_EarthquakeTimeHistories}, through time-domain wavelet superposition \citep{AlAtik_2010_ImprovedMethodNonstationary}, or through Hilbert-Huang decomposition \citep{Ni_2011_ApplicationHilberthuangTransform}. These methods are practical, but they depend on seed-record selection and may alter the nonstationary character of the original motion \citep{Huang_2017_EnergycompatibleSpectrumcompatibleECSC}. Stochastic ground-motion simulation describes seismic excitation through stochastic processes and their parameters. It has been combined with the direct probability integral method for reliability assessment of high concrete-faced rockfill dams under non-stationary ground motions \citep{Lu_2024_SimulationNonstationaryGround} and for dynamic reliability analysis of structures subjected to mainshock--aftershock sequences \citep{Pang_2023_StochasticMainshockAftershock}. Physics-based broadband simulation \citep{MartinMai_2003_HybridMethodCalculating,Graves_2010_BroadbandGroundmotionSimulation,Rodgers_2020_Regionalscale3DGroundmotion} gives a mechanistic route to waveforms, although velocity-model resolution and computational cost limit routine high-frequency use. Stochastic finite-fault and hybrid physics-stochastic methods \citep{Graves_2010_BroadbandGroundmotionSimulation, Boore_2003_SimulationGroundMotion} extend coverage but rely on randomised phases and have difficulty preserving observed inter-frequency amplitude correlations. Engineering applications still lack an efficient probabilistic model that directly samples scenario-conditioned three-component waveforms.

Deep generative models offer a data-driven path from seismological parameters to waveforms. Conditional and Wasserstein generative adversarial networks have been used to synthesise three-component time histories under source, path, and site conditions \citep{Florez_2022_DatadrivenSynthesisBroadband,Esfahani_2023_TFCGANNonstationaryGroundmotion,Shi_2024_BroadbandGroundmotionSynthesis,Matsumoto_2023_FundamentalStudyProbabilistica,Yamaguchi_2024_SitespecificGroundmotionWaveform,Lin_2025_NearfaultGroundMotion,Chen_2025_GroundMotionSimulation}. Variational and dynamic-VAE models represent waveform or spectral variability through latent variables \citep{Ren_2024_LearningPhysicsUnveiling,Ren_2026_LearningEarthquakeGround}. Denoising-diffusion models have been applied in time-domain, latent, and spectrogram representations \citep{Bergmeister_2024_HighResolutionSeismic,Huang_2025_GroundmotionGenerationsUsing,Bi_2025_AdvancingDatadrivenBroadbanda,Jung_2025_BroadbandGroundMotion}. Matsumoto et al. \citep{Matsumoto_2026_WaveformBasedProbabilisticSeismic} incorporated a generative ground motion model into waveform-level PSHA, forming hazard-consistent waveform sets through Monte Carlo sampling, and examined the consistency between the intensity measure hazard derived from these waveforms and conventional PSHA \citep{Kumar_2024_ProbabilisticSeismicHazard,Kumar_2025_EstimationEarthquakeRecurrence}. In that framework, the conditional waveform model generates three-component ground motions from scenarios sampled using the seismic source model. 

Computational cost is another constraint when a hazard calculation spans many scenarios and many samples. Diffusion models \citep{Sohl-Dickstein_2015_DeepUnsupervisedLearning,Ho_2020_DenoisingDiffusionProbabilistic} and their design refinements \citep{Karras_2022_ElucidatingDesignSpace,Rombach_2021_HighresolutionImageSynthesis,Peebles_2022_ScalableDiffusionModels} often require iterative sampling. Distillation and consistency-model training reduce this cost \citep{Salimans_2022_ProgressiveDistillationFast,Sauer_2023_AdversarialDiffusionDistillation,Song_2023_ConsistencyModels,Song_2023_ImprovedTechniquesTraining,Lu_2025_SimplifyingStabilizingScaling}. Flow matching provides a related route by learning a velocity field that transports a simple distribution to the data distribution \citep{Lipman_2022_FlowMatchingGenerative,Liu_2022_FlowStraightFast}. MeanFlow and AlphaFlow further target stable average-velocity learning for one-step or few-step generation \citep{Geng_2025_MeanFlowsOnestep,Zhang_2025_AlphaFlowUnderstandingImproving}. For ground-motion applications, the remaining question is whether few-step generation can preserve the scaling, uncertainty, and waveform realism expected from engineering GMMs.

Models conditioned on response spectra, Husid curves, or energy parameters \citep{Ding_2026_WaveletPacketBasedDiffusion} can generate ground motions that satisfy specified engineering targets, but these targets usually need to be provided in advance by another model based on magnitude, distance, and site conditions. Directly learning absolute amplitude and waveform shape in one high-dimensional target is also limited by sparse and imbalanced data. The model must cover several orders of magnitude in amplitude while retaining nonstationary waveform structure. To address these issues, we propose WaveFlowGMM, which connects scenario conditioning to waveform generation through a joint probability distribution of three-component PGA. The amplitude stage uses physics-informed symbolic learning (PISL) \citep{Chen_2024_PhysicsSymbolicLearner,Liu_2025_PhysicsguidedSymbolicNeural,Liu_2025_HybridSymbolicLearning} to estimate component-wise PGA medians and performs joint amplitude sampling with a full cross-component residual covariance. The waveform stage uses AlphaFlow \citep{Zhang_2025_AlphaFlowUnderstandingImproving} to generate normalised three-component wavelet-packet coefficients conditioned on the scenario variables and sampled PGA. Inverse transformation and amplitude restoration then produce the acceleration histories.

This framework models the low-dimensional amplitude distribution separately from the high-dimensional conditional waveform distribution, allowing scenario-dependent amplitude scaling to be assessed independently of waveform shape. Joint sampling of the three-component PGA carries amplitude dependence into the final records, while few-step AlphaFlow reduces the computational cost of repeated waveform generation. The main contribution is a three-component probabilistic PGA interface coupled to conditional waveform generation. We evaluate the resulting model on held-out earthquake events.

\section{Methodology}\label{sec:methods}

\subsection{Two-stage conditional framework}

Conditional generation of three-component ground-motion histories must control the intensity level and the normalised waveform shape at the same time. The intensity level sets the overall amplitude scale of a record, whereas the normalised waveform shape carries the spectral content, duration, envelope, and inter-component relations. A direct model of the conditional distribution from $\mathbf{x}$ to the three-component acceleration history must span several orders of magnitude in amplitude and a high-dimensional waveform-shape space at once. We therefore factorise the conditional generation problem into two coupled probabilistic models. The first-stage PGA amplitude model predicts the joint distribution of the three-component PGA from the scenario condition $\mathbf{x}$, and the second-stage AlphaFlow generator produces the PGA-normalised three-component waveform given the sampled PGA anchor.
The main symbols introduced in Sections 2.2--2.4 are summarised in Appendix~\ref{app:notation}.

\begin{figure*}[!htbp]
\centering
\includegraphics[width=0.85\textwidth]{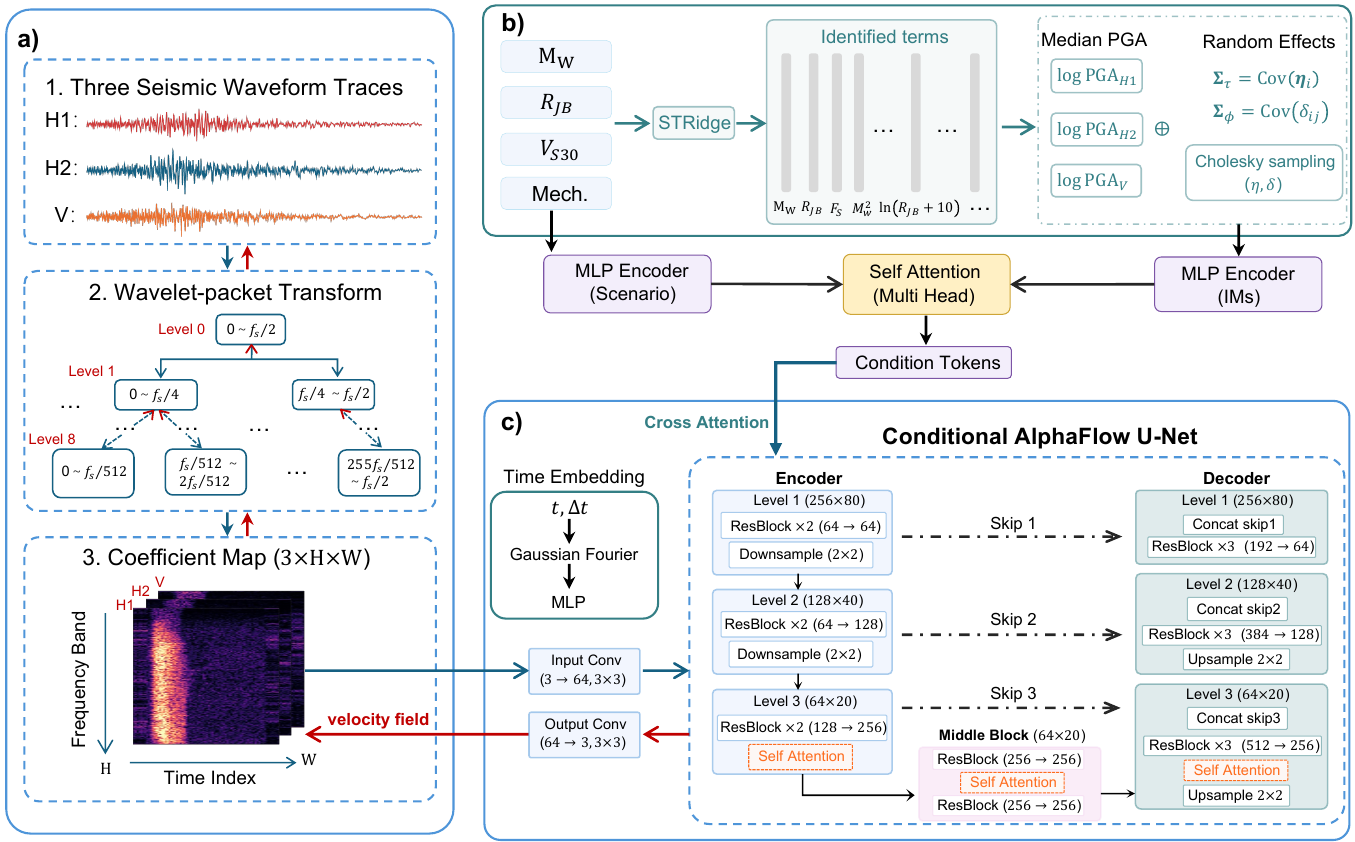}
\caption{Two-stage WaveFlowGMM framework for three-component ground-motion generation. (a) Wavelet-packet coefficient representation; red arrows denote the inverse-transform reconstruction path. (b) PISL amplitude stage, giving the three-component PGA medians and sampling amplitude anchors from the between-event covariance $\Sigma_\tau$ and within-event covariance $\Sigma_\phi$. (c) Conditional U-Net waveform generator, in which scenario and PGA tokens modulate the velocity field through cross-attention.}
\label{fig:framework}
\end{figure*}

Specifically, WaveFlowGMM writes the marginal waveform distribution as an integral over the PGA auxiliary variable $\mathbf{q}$ (Figure~\ref{fig:framework}),
\begin{equation}\label{eq:factor}
p(\mathbf{a}^{1:3}_{0:N}\mid \mathbf{x}) = \int p_{\theta}(\mathbf{a}^{1:3}_{0:N}\mid \mathbf{x}, \mathbf{r}(\mathbf{q}))\,p_{\psi}(\mathbf{q}\mid \mathbf{x})\,\mathrm{d}\mathbf{q}.
\end{equation}
Here $\mathbf{a}^{1:3}_{0:N}$ is the three-component acceleration history and $\mathbf{x}$ is the scenario condition vector, comprising moment magnitude $M_w$, Joyner-Boore distance $R_{JB}$, time-averaged shear-wave velocity over the upper 30\,m $V_{S30}$, and faulting mechanism. The auxiliary variable $\mathbf{q}$ collects the base-10 logarithms of the three-component PGA,
\begin{equation}
\mathbf{q} = [\log_{10}\mathrm{PGA}_{H1},\;\log_{10}\mathrm{PGA}_{H2},\;\log_{10}\mathrm{PGA}_{V}]^\top,
\end{equation}
where H1 and H2 denote the two orthogonal horizontal components and V denotes the vertical component. To pass the three-component amplitude information to AlphaFlow in a more compact form, the sampled PGA triple is further transformed into
\begin{equation}\label{eq:flow_condition}
\begin{split}
\mathbf{r}(\mathbf{q}) = [\,&\log_{10}\mathrm{PGA}_{GM},\;\log_{10}(\mathrm{PGA}_{H1}/\mathrm{PGA}_{H2}),\\
&\log_{10}(\mathrm{PGA}_{V}/\mathrm{PGA}_{GM})\,]^\top,
\end{split}
\end{equation}
where $\mathrm{PGA}_{GM}=\sqrt{\mathrm{PGA}_{H1}\cdot\mathrm{PGA}_{H2}}$ is the geometric mean of the two horizontal components. The first entry controls the overall horizontal amplitude level, the second describes the relative contrast between the two horizontal components, and the third describes the ratio of the vertical component to the horizontal geometric mean.

AlphaFlow models the PGA-normalised waveform coefficients. During training, each component is normalised by its observed PGA before the wavelet-packet transform, $\tilde{a}_c[n] = a_c[n]/\mathrm{PGA}_c$. At inference, a PGA triple is first drawn from $p_{\psi}(\mathbf{q}\mid\mathbf{x})$, and $\mathbf{r}(\mathbf{q})$ passes the amplitude level and the three-component amplitude ratios to AlphaFlow. The generator $p_{\theta}$ then produces the normalised waveform conditioned on $\mathbf{x}$ and $\mathbf{r}(\mathbf{q})$. After inverse wavelet packet transformation, each component is first normalised by its own maximum absolute value, and then each is multiplied by the sampled PGA of the amplitude model in the first stage. This design lets the low-dimensional probabilistic PGA model describe the amplitude scale and the three-component amplitude dependence, while AlphaFlow learns the spectral shape, duration, envelope, and inter-component relations of the conditional normalised waveform.

\subsection{PISL PGA amplitude model}\label{ssec:pga_aux}

Given the scenario condition $\mathbf{x}$, the amplitude stage describes the conditional distribution $p_{\psi}(\mathbf{q}\mid\mathbf{x})$ of the three-component PGA target $\mathbf{q}$. This distribution has two parts: the conditional mean $\boldsymbol{\mu}(\mathbf{x})$ sets the median PGA scaling, and the between-event and within-event covariance set the joint three-component dispersion. The conditional mean is a sparse analytic expression obtained with physics-informed symbolic learning (PISL). Here, physics-informed refers to empirical ground-motion knowledge used to design the candidate functions, without imposing governing physical constraints. The PISL selects a small number of interpretable terms from a candidate library of magnitude, distance, site, and faulting-mechanism terms. STRidge balances fitting error against expression complexity \citep{Chen_2024_PhysicsSymbolicLearner,Liu_2025_HybridSymbolicLearning}. The site term is a saturated $V_{S30}$ ratio, 
\begin{equation}
F_S = \frac{\min(V_{S30}, V_C)}{V_{\mathrm{REF}}},\qquad V_{\mathrm{REF}}=760~\mathrm{m/s},\quad V_C=925~\mathrm{m/s},
\end{equation}
which limits PGA site amplification at high-$V_{S30}$ sites. For each component $d\in\{H1,H2,V\}$, PISL searches the candidate library $\Theta(\mathbf{x}) = \{M_w,\allowbreak R_{JB},\allowbreak F_S,\allowbreak \ln F_S,\allowbreak M_w^2,\allowbreak R_{JB}^2,\allowbreak F_S^2,\allowbreak \ln(R_{JB}+10),\allowbreak M_w\ln(R_{JB}+10),\allowbreak F_M\}$ independently, where $F_M$ encodes the faulting mechanism. The three components retain the same six-term structure with separately fitted coefficients,
\begin{equation}
\begin{aligned}
\mu_d(\mathbf{x}) ={}& \beta_{d,1}M_w + \beta_{d,2}R_{JB} + \beta_{d,3}F_S \\
&+ \beta_{d,4}\ln(R_{JB}+10) \\
&+ \beta_{d,5}M_w\ln(R_{JB}+10) + \beta_{d,6}M_w^2,
\end{aligned}
\end{equation}
where $ \mu_d $ is the conditional mean of $ \log_{10}\mathrm{PGA}_d $, and the three-component fitting coefficients are all determined by the training set (Table~\ref{tab:pisl_coef}).

These terms reference the magnitude scaling, distance attenuation, magnitude-distance coupling, and site response forms commonly used in empirical GMMs, with the aim of restricting the symbolic search to a function space with clear engineering meaning.
The $M_w$ and $M_w^2$ terms allow nonlinear, saturation-like magnitude scaling, consistent with the magnitude-saturation prior discussed by Chen et al. \citep{Chen_2024_PhysicsSymbolicLearner}. However, the model does not impose magnitude saturation or global monotonicity as hard constraints, and the physical plausibility of the resulting expressions comes from the prior design of the candidate functions and their scaling behaviour within the data range rather than from governing equations. The term $\ln(R_{JB}+10)$ describes geometrical spreading with distance and near-source regularisation. The interaction $M_w\ln(R_{JB}+10)$ lets larger events attenuate more slowly, and $F_S$ describes the first-order site effect. Compared with a fully black-box amplitude network, this mean model is easier to diagnose and provides a stable residual baseline for the subsequent covariance sampling.

\begin{table}[!htbp]
\centering
\caption{Coefficients of the PISL three-component PGA median model.}
\label{tab:pisl_coef}
\begin{tabular}{lccc}
\toprule
Term & H1 & H2 & V \\
\midrule
$M_w$                & 1.334 & 1.323 & 1.252 \\
$R_{JB}$             & $-0.00264$ & $-0.00262$ & $-0.00245$ \\
$F_S$                & $-0.483$ & $-0.468$ & $-0.407$ \\
$\ln(R_{JB}+10)$     & $-1.132$ & $-1.126$ & $-1.075$ \\
$M_w\ln(R_{JB}+10)$  & $0.1027$ & $0.1019$ & $0.0840$ \\
$M_w^2$              & $-0.108$ & $-0.107$ & $-0.0974$ \\
\bottomrule
\end{tabular}
\end{table}

The dispersion of the PGA model is estimated from training-set residuals. For record $j$ from event $i$, the three-component PGA residual is
\begin{equation}
\boldsymbol{\zeta}_{ij} = \mathbf{q}_{ij}^{\star} - \boldsymbol{\mu}(\mathbf{x}_{ij}),
\end{equation}
where $\boldsymbol{\mu}(\mathbf{x})=[\mu_{H1}(\mathbf{x}),\,\mu_{H2}(\mathbf{x}),\,\mu_{V}(\mathbf{x})]^\top$. Following the random-effects residual decomposition of Abrahamson and Youngs (1992) \citep{Abrahamson_1992_StableAlgorithmRegression}, this residual is represented as the sum of an event term shared by records from the same earthquake and a within-event term that captures the remaining station-to-station variability,
\begin{equation}
\boldsymbol{\zeta}_{ij}=\boldsymbol{\eta}_i+\boldsymbol{\delta}_{ij}.
\end{equation}
Here $\boldsymbol{\eta}_i$ denotes the systematic event term for earthquake $i$, and $\boldsymbol{\delta}_{ij}$ denotes the within-event residual after the event term has been removed. We apply this scalar random-effects idea to the H1, H2, and V PGA residual vector and estimate full covariance matrices for the between-event and within-event terms,
\begin{equation}
\boldsymbol{\Sigma}_{\tau} = \mathrm{Cov}(\boldsymbol{\eta}_i),\qquad \boldsymbol{\Sigma}_{\phi} = \mathrm{Cov}(\boldsymbol{\delta}_{ij}),
\end{equation}
both of which are $3\times 3$ matrices. Because earthquakes contribute unequal numbers of records, the covariance estimation accounts for unbalanced event sizes rather than using the naive sample covariance of event-mean residuals. The resulting $\boldsymbol{\Sigma}_{\tau}$ and $\boldsymbol{\Sigma}_{\phi}$ describe the joint H1, H2, and V variability at the between-event and within-event levels. With a Gaussian residual approximation, the PGA amplitude model is
\begin{equation}
p^{\mathrm{PISL}}(\mathbf{q}\mid\mathbf{x}) = \mathcal{N}\!\left(\mathbf{q};\boldsymbol{\mu}(\mathbf{x}),\,\boldsymbol{\Sigma}_{\tau}+\boldsymbol{\Sigma}_{\phi}\right).
\end{equation}
At inference, the PGA vector is sampled as
\begin{equation}
\hat{\mathbf{q}} = \boldsymbol{\mu}(\mathbf{x}) + \boldsymbol{\eta} + \boldsymbol{\delta},\qquad \boldsymbol{\eta}\sim\mathcal{N}(\mathbf{0},\boldsymbol{\Sigma}_{\tau}),\quad \boldsymbol{\delta}\sim\mathcal{N}(\mathbf{0},\boldsymbol{\Sigma}_{\phi}),
\end{equation}
where $\boldsymbol{\eta}$ and $\boldsymbol{\delta}$ are drawn independently with Cholesky factors of the two covariance matrices.

In this study, $\boldsymbol{\Sigma}_{\tau}$ and $\boldsymbol{\Sigma}_{\phi}$ are global constant matrices estimated from the full training set. They approximate three-component PGA residual dependence at the dataset level and remain fixed across magnitude, distance, site condition, and faulting mechanism. They only describe the joint residuals of the three-component PGA scalars, and do not parameterise the phase, envelope, or time-frequency relationships of the waveforms; the latter type of cross-component dependence is learned by the three-channel AlphaFlow under given PGA conditions.

Because the matrices are full rather than diagonal, the sampled H1, H2, and V values preserve the observed cross-component dependence. This retains the smooth and interpretable PISL mean while giving the three-component PGA vector a probabilistic dispersion consistent with the between-event and within-event residual structure.

\subsection{AlphaFlow training objective in wavelet-packet space}\label{ssec:alphaflow}
AlphaFlow models in the three-component wavelet packet coefficient space after PGA normalisation. This representation preserves a reconstructible time-frequency structure, enabling the generator to learn the waveform envelope, frequency-band energy distribution, and cross-component coupling in a compact coefficient space.

Each component's 16,384-point time history is decomposed using sym8 eight-level wavelet packet decomposition \citep{Ding_2026_WaveletPacketBasedDiffusion,Daubechies_1988_OrthonormalBasesCompactly}, with symmetric boundary extension. The eighth level contains 256 frequency band nodes, each with a length of 78, so the single-component coefficient matrix is $256\times78$. To adapt to the downsampling scale of U-Net, the coefficient matrix is padded with two columns of zeros along the length direction, and the complete size of the three-component network input is $3\times256\times80$. Before the inverse transform, the two zero-padded columns are removed from the network output. The remaining $3\times256\times78$ coefficients are then transformed back using the same wavelet basis and boundary rules. 

Given scenario condition $\mathbf{x}$ and PGA condition $\mathbf{r}$, let $\mathbf{w}_0$ denote the wavelet-packet coefficients of the PGA-normalised waveform and let $\mathbf{z}\sim\mathcal{N}(\mathbf{0},\mathbf{I})$ denote prior noise. Flow matching \citep{Lipman_2022_FlowMatchingGenerative} defines a linear training path between $\mathbf{w}_0$ and $\mathbf{z}$,
\begin{equation}
\mathbf{w}_t=(1-t)\mathbf{w}_0+t\mathbf{z},\qquad t\in[0,1],
\end{equation}
with target velocity
\begin{equation}
\mathbf{v}_t=\mathbf{z}-\mathbf{w}_0.
\end{equation}
The neural network $\mathbf{u}_{\theta}(\mathbf{w}_t,t,\Delta t;\mathbf{x},\mathbf{r})$ is additionally conditioned on the step size $\Delta t$ and learns the conditional average velocity over the interval $[t-\Delta t,\,t]$, reducing to the instantaneous velocity field $\mathbf{v}_t$ above as $\Delta t\to 0$. Generation starts from $\mathbf{w}_1\sim\mathcal{N}(\mathbf{0},\mathbf{I})$ and integrates backward to $t=0$. With Euler updates, one step is
\begin{equation}\label{eq:euler_step}
\mathbf{w}_{t-\Delta t}=\mathbf{w}_t-\Delta t\cdot
\mathbf{u}_{\theta}(\mathbf{w}_t,t,\Delta t;\mathbf{x},\hat{\mathbf{r}}),
\end{equation}
where $\hat{\mathbf{r}}$ is assembled from the sampled three-component PGA.

Standard flow matching learns an instantaneous velocity field along a continuous path. This objective is well matched to fine-step ODE integration; when the sampling budget is compressed to one or two steps, a single network evaluation must represent average transport over a long time interval, and the local velocity along the linear path can amplify discretisation error. In three-component wavelet-packet coefficients, this error can appear as envelope stretching, band-energy shifts, or duration distortion. We use the AlphaFlow $\alpha$ schedule \citep{Zhang_2025_AlphaFlowUnderstandingImproving}, which places trajectory flow matching (TFM), Shortcut Models \citep{Frans_2024_OneStepDiffusion}, and MeanFlow \citep{Geng_2025_MeanFlowsOnestep} in one training framework. During training, $\alpha$ is annealed with relative progress $s$,
\begin{equation}
\begin{aligned}
\alpha(s) &= \alpha_{\mathrm{end}} + (\alpha_{\mathrm{init}} - \alpha_{\mathrm{end}})\,
\mathrm{sigmoid}\!\bigl(-\gamma(\tilde s-\tfrac{1}{2})\bigr),\\
\tilde s &=\mathrm{clamp}\!\left(\frac{s-s_{\mathrm{start}}}{s_{\mathrm{end}}-s_{\mathrm{start}}},0,1\right).
\end{aligned}
\end{equation}
Here $s\in[0,1]$ is the fraction of completed training steps, and $s_{\mathrm{start}}$ and $s_{\mathrm{end}}$ are the progress thresholds that bound the $\alpha$ anneal. Each mini-batch is divided by samples into two sub-batches. Approximately half of the samples compute the TFM objective, and the other half compute the MeanFlow objective. The two sub-batches share the same U-Net and condition encoder, with different training targets. Flow time is obtained from a log-normal noise scale,
\begin{equation}
t=\frac{\sigma_n}{1+\sigma_n},\qquad \ln\sigma_n\sim\mathcal{N}(-0.4,1.0),
\end{equation}
following the EDM noise-scale parametrisation \citep{Karras_2022_ElucidatingDesignSpace}.

The two objectives are trained jointly with an adaptively normalised squared-error loss,
\begin{equation}
\mathcal{L}_{\mathrm{AF}} =
\mathbb{E}_{\mathbf{w}_0,\mathbf{z},t,b}
\left[
\frac{\lambda_{\alpha}}{\|\mathbf{e}\|^2_{\mathrm{sg}}+\varepsilon_0}
\|\mathbf{e}\|^2
\right],\qquad
\mathbf{e}=\mathbf{u}_{\theta}-\mathbf{v}^{\mathrm{tgt}}_t,
\end{equation}
where $b\in\{\mathrm{TFM},\mathrm{MF}\}$ denotes the training branch, $\lambda_{\alpha}=1$ for the TFM branch, $\lambda_{\alpha}=\alpha$ for the MeanFlow branch, $\|\cdot\|_{\mathrm{sg}}$ denotes the stop-gradient term, and $\varepsilon_0=10^{-3}$ stabilises the normalisation weight.

The target velocity $\mathbf{v}^{\mathrm{tgt}}_t$ depends on the training branch and equals the average velocity over $[t-\Delta t,\,t]$, $\mathbf{u}(\mathbf{w}_t,t,\Delta t)=\frac{1}{\Delta t}\int_{t-\Delta t}^{t}\mathbf{v}_s\,\mathrm{d}s$. The TFM branch sets $\Delta t=0$, so the target reduces to the instantaneous velocity $\mathbf{v}^{\mathrm{tgt}}_t=\mathbf{v}_t=\mathbf{z}-\mathbf{w}_0$; the MeanFlow branch sets $\Delta t>0$, where the target follows the MeanFlow identity
\begin{equation}\label{eq:meanvel}
\mathbf{v}^{\mathrm{tgt}}_t=\mathbf{v}_t-\Delta t\,\frac{\mathrm{d}}{\mathrm{d} t}\,\mathbf{u}_{\theta}(\mathbf{w}_t,t,\Delta t;\mathbf{x},\mathbf{r}),
\end{equation}
with the total derivative $\mathrm{d}\mathbf{u}_{\theta}/\mathrm{d} t$ along the trajectory computed by a single Jacobian-vector product; the $\alpha$ schedule interpolates this target between the continuous identity above and a Shortcut-style discrete bootstrap. A single network evaluation in few-step sampling therefore represents the average velocity over a given step rather than the instantaneous velocity.

During training, AlphaFlow is conditioned on $\mathbf{r}^{\star}$ computed from the observed PGA. We use $\alpha_{\mathrm{init}}=1.0$, $\alpha_{\mathrm{end}}=0.4$, $\gamma=10$, $s_{\mathrm{start}}=0.25$, and $s_{\mathrm{end}}=0.70$. Training uses the Adam optimiser with an initial learning rate of $1\times10^{-4}$, cosine annealing, a batch size of 64, and 200 epochs.

As the training objective gradually transitions from Trajectory Flow Matching (TFM) to the average velocity field, the sampling quality of AlphaFlow exhibits a non-monotonic variation with the number of neural function evaluations (NFE). Therefore, a larger NFE is not necessarily better \citep{Song_2023_ConsistencyModels, Geng_2025_MeanFlowsOnestep}. The main results are reported at two NFE. Section~\ref{ssec:efficiency} presents the sensitivity analysis for the tested sampling budgets.

\subsection{Conditional encoding, U-Net architecture, and sampling}\label{ssec:network}

The condition encoder represents physical scenario variables as structured condition tokens. Each token corresponds to one physical factor, allowing cross-attention to read source, path, site, and amplitude-anchor information separately. Three tokens encode the scenario: the source token contains $M_w$ and faulting mechanism, the distance token contains $R_{JB}$, and the site token contains $V_{S30}$. Continuous variables are embedded with small multilayer perceptrons. Faulting mechanism is treated as a categorical variable, mapped by a trainable embedding table, mixed with the $M_w$ field in the source encoder, and pooled into the source token. A shallow Transformer encoder \citep{Vaswani_2017_AttentionAllYou} then provides light interaction among the scenario tokens. The PGA condition $\mathbf{r}$ forms a separate amplitude token containing the horizontal geometric-mean amplitude, the relative ratio of the two horizontal components, and the vertical-to-horizontal ratio. The final condition sequence is $[\mathbf{c}_{\mathrm{src}},\mathbf{c}_{\mathrm{dist}},\mathbf{c}_{\mathrm{site}},\mathbf{c}_{\mathrm{amp}}]$, where the first three tokens come from scenario variables and the amplitude token comes from the PISL PGA stage.

The generator backbone is a two-dimensional convolutional U-Net \citep{Ronneberger_2015_UnetConvolutionalNetworks} because the wavelet-packet coefficient array has an explicit time-frequency grid structure. The input is the stacked three-component coefficient array, and the output is a conditional velocity field with the same shape. The network uses 64 base channels, channel multipliers of 1, 2, and 4, and two residual blocks at each scale. Convolutions handle local time-frequency patterns, while attention handles long-range band coupling. Self-attention is placed at the four-fold downsampling scale and in the U-Net bottleneck to model long-range time-frequency coupling. Conditional cross-attention is placed at the full, two-fold, and four-fold resolutions and in the bottleneck, allowing scenario and PGA tokens to modulate the velocity field across multiple feature resolutions. The time step is embedded with Gaussian Fourier features followed by a two-layer MLP. Residual blocks use scale-shift modulation to combine time and condition information. Condition tokens are linearly projected into attention keys and values, while the current U-Net features provide the queries.

Algorithm~\ref{alg:waveflowgmm_sampling} summarises inference. Rescale normalises each component by its peak absolute amplitude and then restores the sampled PGA.

\begin{algorithm}[!htbp]
\caption{WaveFlowGMM inference}
\label{alg:waveflowgmm_sampling}
\begin{algorithmic}[1]
\Require scenario condition $\mathbf{x}$, PGA amplitude model $p_{\psi}$, AlphaFlow average-velocity field $\mathbf{u}_{\theta}$, and AlphaFlow step count $K$
\Ensure Three-component acceleration history $\hat{\mathbf{a}}^{1:3}_{0:N}$
\State $\hat{\mathbf{q}}\sim p_{\psi}(\mathbf{q}\mid\mathbf{x})$
\State $\hat{\mathbf{r}}\gets\mathbf{r}(\hat{\mathbf{q}})$
\State $\mathbf{w}\sim\mathcal{N}(\mathbf{0},\mathbf{I})$, \quad $\Delta t\gets 1/K$
\For{$k=1,\ldots,K$}
\State $t_k\gets 1-(k-1)\Delta t$
\State $\mathbf{w}\gets \mathbf{w}-\Delta t\,\mathbf{u}_{\theta}(\mathbf{w},t_k,\Delta t;\mathbf{x},\hat{\mathbf{r}})$
\EndFor
\State $\tilde{\mathbf{a}}^{1:3}_{0:N}\gets \mathrm{IWPT}_{\mathrm{sym8},8}(\mathbf{w})$
\State $\hat{\mathbf{a}}^{1:3}_{0:N}\gets\mathrm{Rescale}(\tilde{\mathbf{a}}^{1:3}_{0:N},\hat{\mathbf{q}})$
\State \Return $\hat{\mathbf{a}}^{1:3}_{0:N}$
\end{algorithmic}
\end{algorithm}

\section{Strong-Motion Data}\label{sec:data}

The training and test sets are drawn from the NGA-West2 strong-motion database \citep{Ancheta_2014_NGAWest2Database}. Record selection follows the data-screening logic used by the NGA-West2 empirical models \citep{Abrahamson_2014_SummaryASK14Ground,Boore_2014_NGAWest2EquationsPredicting,Campbell_2014_NGAWest2GroundMotion,Chiou_2014_UpdateChiouYoungs}, giving the generator and the reference empirical models a similar population of free-field shallow crustal recordings in active tectonic regions. The screening was applied to the public updated flatfile and the corresponding three-component time histories using the following criteria:

\begin{itemize}
\item Retain shallow crustal earthquakes in active tectonic regions and remove events unsuitable for active-crustal regression, including non-active-crustal events, selected aftershock sequences, events with questionable source depths, and events recorded only by special arrays.
\item Retain free-field station recordings and exclude those that may be affected by soil-structure interaction or installation conditions.
\item Remove records flagged as questionable, including samples with problematic quality or spectral-quality flags and records marked as late S-trigger cases.
\item Require finite key metadata for conditional modelling, including moment magnitude $M_w$, Joyner-Boore distance $R_{JB}$, $V_{S30}$, and faulting mechanism.
\item Require three acceleration components under the same record sequence number.
\item Apply year-, region-, and magnitude-dependent $R_{JB}$ censoring distances to reduce far-distance triggering bias, and require at least three retained records per earthquake after screening.
\end{itemize}

These filters give 15{,}615 three-component records from 358 earthquakes. Following Boore et al. \citep{Boore_2014_NGAWest2EquationsPredicting}, period-dependent PSA statistics use a record only when its usable frequency band reaches the oscillator period. This rule is applied to period-dependent residuals, scatter points, and dispersion estimates in the Results, not as a whole-record exclusion criterion.

\begin{figure*}[!htbp]
\centering
\includegraphics[width=0.8\textwidth]{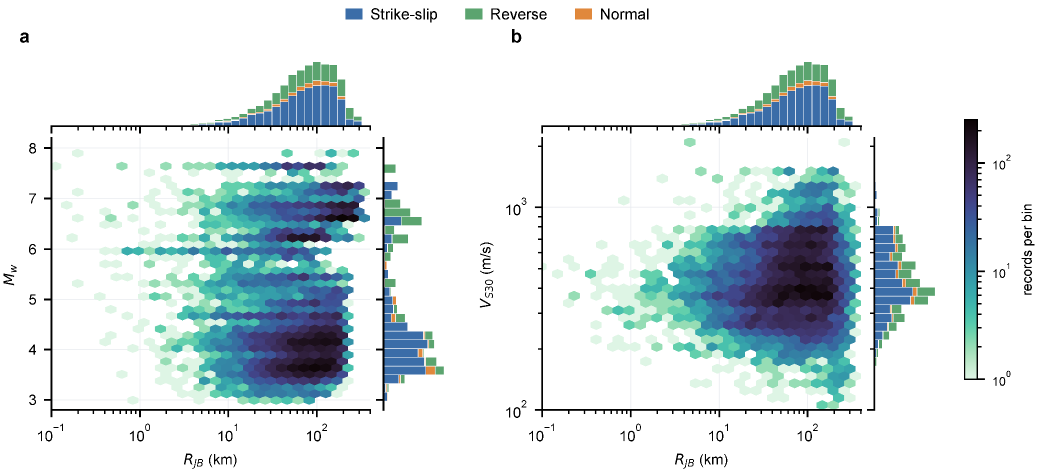}
\caption{Distributions of magnitude, distance, and site velocity in the NGA-West2 training and test sets.}
\label{fig:data_dist}
\end{figure*}

The dataset is split by earthquake rather than by record. With a fixed random seed, whole earthquakes are assigned to the training or test partition in an 85:15 ratio. This event-level split evaluates the model on earthquakes that were unseen during training and prevents records from the same earthquake from appearing in both partitions. The resulting training set contains 12{,}472 records from 304 events, and the event-level holdout set contains 3{,}143 records from 54 events (Figure~\ref{fig:data_dist}).

Table~\ref{tab:data_range} reports the ranges and distributions of the conditioning variables and PGA in the training and evaluation sets. Continuous variables are summarised by their median and minimum-to-maximum range, while faulting mechanisms are reported as event-level proportions.

\begin{table*}[!htbp]
\centering

\caption{Conditioning-variable and PGA ranges in the training and evaluation sets.}
\label{tab:data_range}
\renewcommand{\arraystretch}{1.15}
\begin{tabularx}{\textwidth}{@{}l>{\centering\arraybackslash}X>{\centering\arraybackslash}X@{}}
\toprule
Metric & Training set & Evaluation set \\
\midrule
$M_w$ & 4.06 [2.99--7.90] & 4.42 [3.25--7.28] \\
$R_{JB}$ (km) & 71.94 [0--337.83] & 96.68 [0--344.91] \\
$V_{S30}$ (m/s) & 417.99 [89.32--2{,}100] & 408.20 [89.32--1{,}525.85] \\
PGA$_{H1}$ (cm/s$^2$) & 2.29 [0.0015--1{,}743.82] & 6.49 [0.0020--1{,}707.41] \\
PGA$_{H2}$ (cm/s$^2$) & 2.23 [0.0021--2{,}419.09] & 6.46 [0.0024--1{,}299.86] \\
PGA$_{V}$ (cm/s$^2$) & 1.49 [0.0017--3{,}763.53] & 3.70 [0.0021--1{,}857.07] \\
Faulting mechanism & SS: 64.5\%; NS: 14.1\%; RS: 21.4\% & SS: 64.8\%; NS: 9.3\%; RS: 25.9\% \\
\bottomrule
\end{tabularx}
\vspace{2pt}
\parbox{\textwidth}{\footnotesize\textit{Note:} Continuous variables are reported as median [minimum--maximum]. $M_w$ and faulting mechanism are summarised by event; $R_{JB}$, $V_{S30}$, and PGA are summarised by record.}
\end{table*}

Waveform preprocessing maps all records into a common generation target space. Records are resampled to 100\,Hz by linear interpolation and set to 16{,}384 samples. Longer records are cropped, and shorter records are zero-padded to this length. The three components always use the same cropping window and time shift to preserve their relative timing. We further align the records in time. Specifically, all three-component records use a common time window, with the 5\% cumulative-energy time of the three-component resultant acceleration as the alignment anchor. We then refine the alignment by cross-correlating smoothed Hilbert envelopes and searching for the optimal shift within $\pm5$\,s of the anchor. Any portions shifted beyond the original record are zero-padded. 

Beyond the NGA-West2 processing, no additional tapering was applied during training-data preparation, and no dedicated pre-event noise removal was performed. For the waveform comparisons in Section~\ref{ssec:scaling}, both observed and generated records received additional cosine tapers over the first 1\,s and last 2\,s.

The amplitude-stage target is the three-component PGA vector $\mathbf{q} = [\log_{10}\mathrm{PGA}_{H1},\;\log_{10}\mathrm{PGA}_{H2},\;\log_{10}\mathrm{PGA}_{V}]^\top$. The waveform stage uses acceleration histories normalised by component-wise PGA and represented as sym8 level-8 wavelet-packet coefficient arrays. Horizontal response spectra are computed as RotD50 from the two horizontal components \citep{Boore_2014_NGAWest2EquationsPredicting, Boore_2010_OrientationindependentNongeometricmeanMeasures}.

\section{Results}\label{sec:results}

AlphaFlow training was completed on three NVIDIA L20 GPUs using Distributed Data Parallel (DDP) and BF16 automatic mixed precision, with a training batch size of 64, a total of 200 epochs, and each epoch containing 273 training batches. The batch size in the inference stage is 400.

\subsection{Intensity-measure accuracy and PGA component dependence}\label{ssec:ablation_im_aux}

This section compares observed and generated three-component ground motions using engineering intensity measures. All intensity measures (IMs) are recomputed from the complete acceleration histories generated under the test-set scenarios, covering peak, response-spectral, integrated-energy, and duration measures. Table~\ref{tab:im_def} summarises the definitions and three-component accuracy of the twelve IMs, where $\rho$ denotes the Pearson correlation between observed and generated values and bias is defined as $\ln y_{\mathrm{obs}}-\ln y_{\mathrm{gen}}$. The record-level correlations quantify associations between observed intensity measures and those generated under the corresponding scenario conditions. These correlations evaluate scalar intensity measures and do not assess point-by-point reconstruction of individual observed waveforms.

IMs are calculated separately for each component over the full record window. PGA, response spectrum, Arias intensity, CAV, and significant duration are directly computed from the acceleration waveforms restored to physical units. PGV and PGD are computed from the velocity and displacement waveforms obtained by applying a 0.1 Hz, 4th-order zero-phase Butterworth high-pass filter and trapezoidal integration.

Peak and spectral measures show that the generated waveforms reproduce the main amplitude scale and period-dependent response of observed records. PGA, PGV, and peak spectral measures retain high correlations on all three components, and PGD, PSD$_{\max}$, and Housner spectral intensity also reach correlations of about 0.91 or higher (Table~\ref{tab:im_def}). PGV, PGD, and spectral intensity further indicate that the generated histories retain reasonable velocity, displacement, and band-limited energy content after integration and response-spectrum analysis.

\begin{table*}[!htbp]
\centering
\caption{Definitions and three-component end-to-end accuracy of twelve intensity measures.}
\label{tab:im_def}
\renewcommand{\arraystretch}{1.15}
\setlength{\tabcolsep}{4pt}
\begin{tabular}{@{}llll ccc ccc@{}}
\toprule
\multirow{2}{*}{IM} & \multirow{2}{*}{Name} & \multirow{2}{*}{Definition} & \multirow{2}{*}{Unit} & \multicolumn{3}{c}{$\rho$} & \multicolumn{3}{c}{Bias} \\
\cmidrule(lr){5-7}\cmidrule(lr){8-10}
 & & & & H1 & H2 & V & H1 & H2 & V \\
\midrule
\multicolumn{10}{@{}l}{Peak and spectral measures} \\
PGA & Peak acceleration & $\max|\ddot{u}_g(t)|$ & cm/s$^2$ & 0.883 & 0.881 & 0.883 & $-$0.071 & $-$0.076 & $-$0.098 \\
PGV & Peak velocity & $\max|\dot{u}_g(t)|$ & cm/s & 0.902 & 0.900 & 0.904 & $-$0.186 & $-$0.180 & $-$0.274 \\
PGD & Peak displacement & $\max|u_g(t)|$ & cm & 0.916 & 0.913 & 0.916 & $-$0.288 & $-$0.269 & $-$0.314 \\
PSA$_{\max}$ & Peak spectral acceleration & $\max_T S_A(T)$ & cm/s$^2$ & 0.874 & 0.874 & 0.874 & $-$0.016 & $-$0.023 & $-$0.013 \\
PSV$_{\max}$ & Peak spectral velocity & $\max_T S_V(T)$ & cm/s & 0.895 & 0.894 & 0.898 & $-$0.177 & $-$0.169 & $-$0.244 \\
PSD$_{\max}$ & Peak spectral displacement & $\max_T S_D(T)$ & cm & 0.920 & 0.917 & 0.915 & $-$0.322 & $-$0.298 & $-$0.438 \\
\midrule
\multicolumn{10}{@{}l}{Integrated and energy measures} \\
EPA & Effective peak acceleration & $\overline{S_A}(0.1\text{-}0.5\,\mathrm{s})/2.5$ & cm/s$^2$ & 0.883 & 0.883 & 0.887 & $-$0.211 & $-$0.215 & $-$0.281 \\
SI & Housner spectral intensity & $\int_{0.1}^{2.5}S_V(T)\,\mathrm{d}T$ & cm & 0.913 & 0.911 & 0.914 & $-$0.321 & $-$0.302 & $-$0.368 \\
$I_A$ & Arias intensity & $\frac{\pi}{2g}\int_0^{t_f}\ddot{u}_g^2\,\mathrm{d}t$ & cm/s & 0.891 & 0.890 & 0.891 & $-$0.383 & $-$0.385 & $-$0.401 \\
CAV & Cumulative abs. vel. & $\int_0^{t_f}|\ddot{u}_g|\,\mathrm{d}t$ & cm/s & 0.891 & 0.890 & 0.889 & $-$0.249 & $-$0.250 & $-$0.243 \\
$I_c$ & Characteristic int. & $a_{\mathrm{rms}}^{1.5}\,t_d^{0.5}$ & cm$^{1.5}$/s$^{2.5}$ & 0.893 & 0.892 & 0.890 & $-$0.276 & $-$0.281 & $-$0.261 \\
\midrule
\multicolumn{10}{@{}l}{Duration measure} \\
$D_{5\text{-}95}$ & Significant duration & $t(0.95I_A)-t(0.05I_A)$ & s & 0.709 & 0.703 & 0.730 & 0.022 & 0.017 & 0.079 \\
\bottomrule
\end{tabular}
\par\smallskip
{\footnotesize Note: $\ddot{u}_g$ is the ground acceleration, $g$ the gravitational acceleration, $t_f$ the total record duration, $a_{\mathrm{rms}}$ the root-mean-square acceleration, and $t_d$ the duration used (taken here as $D_{5\text{-}95}$).}
\end{table*}

\begin{table*}[!htbp]
\centering
\caption{Cross-component correlations of three-component PGA.}
\label{tab:pisl_pga}
\renewcommand{\arraystretch}{1.15}
\begin{tabular}{@{}lcccccc@{}}
\toprule
 & \multicolumn{3}{c}{Residual correlation} & \multicolumn{3}{c}{Total PGA correlation} \\
\cmidrule(lr){2-4}\cmidrule(lr){5-7}
Source & H1-H2 & H1-V & H2-V & H1-H2 & H1-V & H2-V \\
\midrule
Observed & 0.937 & 0.861 & 0.864 & 0.993 & 0.983 & 0.983 \\
Full covariance & 0.934 & 0.860 & 0.860 & 0.993 & 0.984 & 0.983 \\
Diagonal (independent) & 0 & 0 & 0 & 0.892 & 0.891 & 0.890 \\
\bottomrule
\end{tabular}
\end{table*}

PGA and $\mathrm{PSA}_{\max}$ have biases close to zero, suggesting that the generated waveforms retain peak amplitude and peak spectral response reasonably well. The mean logarithmic residuals for PGV, PGD, SI, Arias intensity, CAV, and $I_c$ are negative, so their generated values have higher geometric means than the observations. PGA rescaling controls each component's peak. Other IMs also depend on the spectrum, envelope, and duration of the normalised waveform. Source and path variability absent from the conditioning variables, low-frequency content, and integration processing may contribute to these biases. The same high-pass filtering and trapezoidal integration are used to calculate PGV and PGD from observed and generated waveforms. The contribution of each factor remains unresolved.

Overestimated PGD and $\mathrm{PSD}_{\max}$ may lead to higher estimates of displacement-sensitive engineering demands. Higher generated Arias intensity, CAV, and $I_c$ may affect cumulative-energy assessments. These IMs should be checked and calibrated for the intended structural-demand or hazard analysis. Duration has relatively low record-level correlation, so its predictive distribution also needs to be evaluated for applications that depend on duration. Response analyses are needed to determine how these IM biases affect specific structures.

The median significant duration $D_{5\text{-}95}$ is close to that of the observed records, but its record-level correlation is lower than those of the amplitude and spectral measures. This lower predictability is expected because significant duration is mainly controlled by the source rupture process and path scattering, which are not included as explicit model inputs. When only magnitude, distance, site condition, and faulting mechanism are specified, record-level duration predictability has an inherent limit. The empirical duration model of Afshari and Stewart \citep{Afshari_2016_PhysicallyParameterizedPrediction}, for example, also predicts $D_{5\text{-}95}$ mainly from magnitude, distance, and site terms and retains substantial aleatory scatter.

Accurate single-component PGA does not by itself ensure a realistic joint amplitude structure across the three waveform components. Full-covariance sampling reproduces the three-component residual correlations in the test set, whereas the diagonal covariance forces residual correlations to zero by construction (Table~\ref{tab:pisl_pga}).

\begingroup After training, the obtained between-event, within-event, and total residual covariance matrices are given below. The matrix row and column order is H1, H2, and V, and all matrices are defined in the $\log_{10}\mathrm{PGA}$ domain:

\begin{equation*}
\boldsymbol{\Sigma}_\tau=\begin{bmatrix}0.0398&0.0396&0.0404\\0.0396&0.0395&0.0403\\0.0404&0.0403&0.0437\end{bmatrix},
\end{equation*}

\begin{equation*}
\boldsymbol{\Sigma}_\phi=\begin{bmatrix}0.1084&0.0998&0.0844\\0.0998&0.1108&0.0854\\0.0844&0.0854&0.0984\end{bmatrix},
\end{equation*}

\begin{equation*}
\boldsymbol{\Sigma}=\boldsymbol{\Sigma}_\tau+\boldsymbol{\Sigma}_\phi=\begin{bmatrix}0.1482&0.1394&0.1248\\0.1394&0.1503&0.1257\\0.1248&0.1257&0.1421\end{bmatrix}.
\end{equation*}

The corresponding correlation matrix is
\begin{equation*}
\mathbf{R}_\tau=\begin{bmatrix}1&0.999&0.969\\0.999&1&0.970\\0.969&0.970&1\end{bmatrix},
\end{equation*}

\begin{equation*}
    \mathbf{R}_\phi=\begin{bmatrix}1&0.911&0.817\\0.911&1&0.818\\0.817&0.818&1\end{bmatrix},
\end{equation*}

\begin{equation*}
    \mathbf{R}=\begin{bmatrix}1&0.934&0.860\\0.934&1&0.860\\0.860&0.860&1\end{bmatrix}.
\end{equation*}

\endgroup

From the full covariance model, the generated samples have H1-H2, H1-V, and H2-V residual correlation coefficients of $0.934$, $0.860$, and $0.860$, respectively, which are close to the test-recorded values of $0.937$, $0.861$, and $0.864$. This indicates that the model can reproduce the overall three-component PGA residual dependence in the current test set.

\begin{figure*}[!htbp]
\centering
\includegraphics[width=0.8\textwidth]{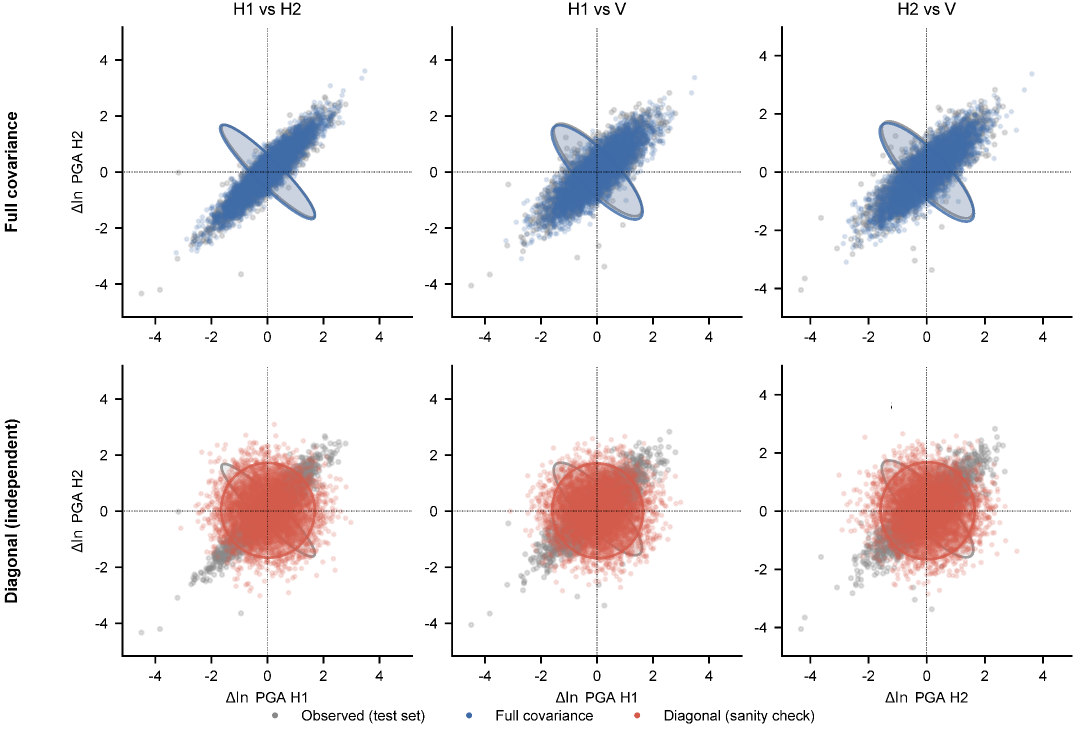}
\caption{Cross-component PGA residual dependence under full and diagonal covariance models.}
\label{fig:pga_cov_ablation}
\end{figure*}

Under diagonal covariance, the total PGA still retains about 0.89 of the component correlation, because the three components share the same conditional mean and vary together with the same magnitude, distance, site conditions, and fault mechanism. However, these correlations are lower than the 0.98--0.99 observed in the records and do not include component-ratio variability from the aleatory residuals. The residual scatter in Figure~\ref{fig:pga_cov_ablation} gives the same conclusion. Full-covariance samples overlap the test-set residual cloud, whereas independent-component sampling cannot reproduce the observed oblique dependence structure. The full covariance therefore extends the PGA amplitude interface from three independent marginal quantities to a three-component random vector and passes the observed cross-component dependence to subsequent waveform generation.

\begin{figure*}[!htbp]
\centering
\includegraphics[width=0.8\textwidth]{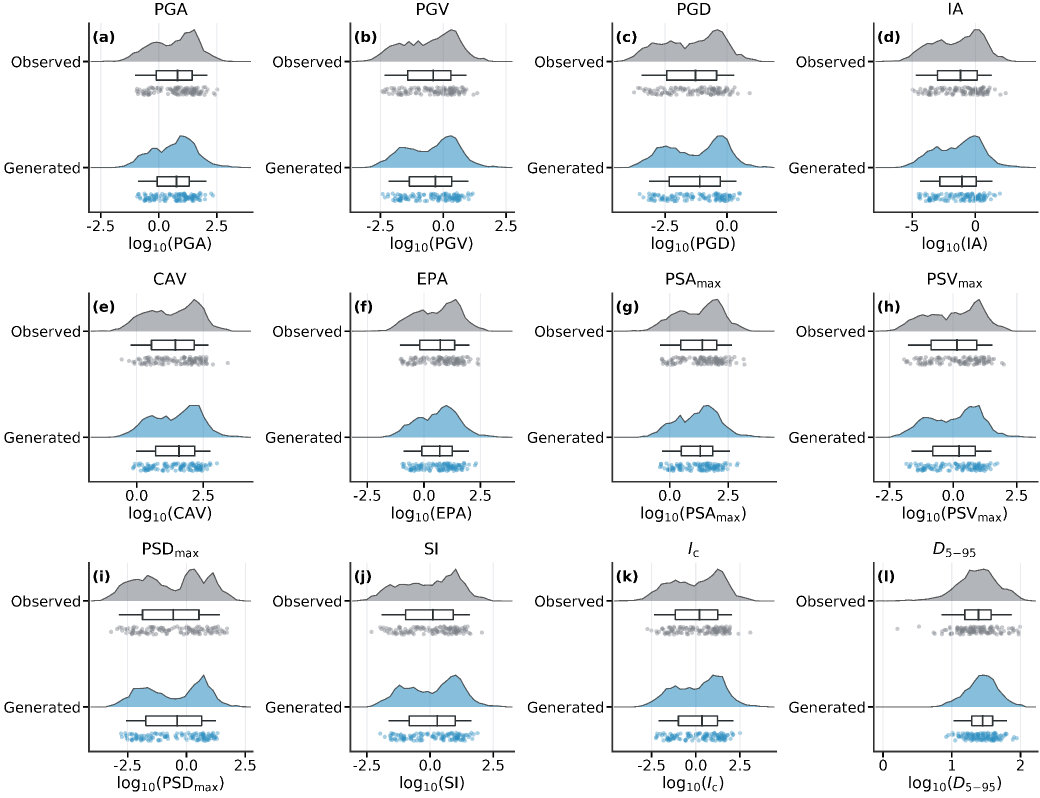}
\caption{Marginal distributions of twelve intensity measures for observed and generated H1 ground motions. Shaded profiles show the distribution shapes. Thin horizontal lines span the 5th to 95th percentiles, boxes span the 25th to 75th percentiles, and vertical lines inside the boxes mark the medians. Dots show a subset of sample values.}
\label{fig:im_density}
\end{figure*}

Figure~\ref{fig:im_density} complements the correlation and bias statistics by showing distributional shapes; owing to length constraints, only the H1 component is shown. The observed and generated H1 IMs overlap over the main amplitude range, and the peak and spectral measures show no tail truncation or mode splitting. Integrated and energy measures shift slightly towards higher generated values, consistent with the negative biases in Table~\ref{tab:im_def}. The $D_{5\text{-}95}$ distribution has a median close to the observed records but remains broad, consistent with its lower record-level correlation.

\subsection{Median scaling for distance, magnitude, and site response}\label{ssec:scaling}

\begin{figure*}[!htbp]
\centering
\includegraphics[width=0.8\textwidth]{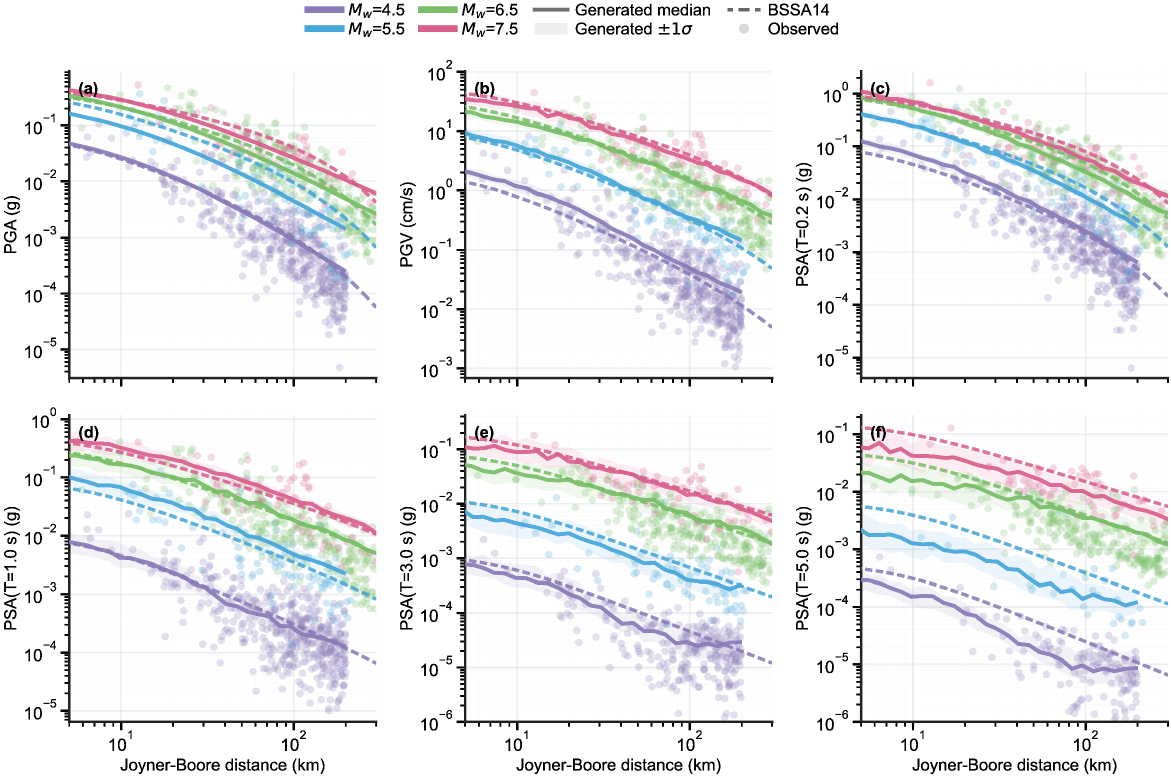}
\caption{Distance attenuation of RotD50 intensity measures. PGA, PGV, and PSA at 0.2, 1, 3, and 5\,s are shown as functions of $R_{JB}$ for strike-slip conditions at $V_{S30}=760$\,m/s.}
\label{fig:distance_scaling}
\end{figure*}

Empirical GMM results sections usually use median-scaling curves to test whether source, path, and site terms are reasonable. Here, $R_{JB}$, $M_w$, and $V_{S30}$ are varied one at a time while the remaining conditions are fixed. We compare the generated-sample medians and 16th-to-84th-percentile bands for RotD50 PGA, PGV, and PSA. The path-attenuation curves show that the generated waveforms retain the basic structure of attenuation with distance and amplitude increase with magnitude. For strike-slip conditions at $V_{S30}=760$\,m/s, the median curves for $M_w=4.5$ to 7.5 decay smoothly from 1 to 300\,km, flatten towards short distances, and preserve the magnitude ordering at all distances (Figure~\ref{fig:distance_scaling}).

Within the data-supported range, PGA, PGV, and short-period PSA show broadly similar attenuation trends to the BSSA14 \citep{Boore_2014_NGAWest2EquationsPredicting} reference curves and stay within the main observation scatter. The 3 and 5\,s data are more scattered, but the generated medians remain within the usable-record scatter after observations are restricted to records usable at each period. In the small-magnitude, far-distance range, the generated medians approach a plateau close to the observed records, a shape that differs from BSSA14. This indicates that in the far-field long-period corner where data coverage is weak, the generator's median is mainly constrained by the distribution of available records and cannot reflect a stable physical long-period scaling relationship. Sparse observations limit the assessment of far-field long-period attenuation in this range.

\begin{figure*}[!htbp]
\centering
\includegraphics[width=0.8\textwidth]{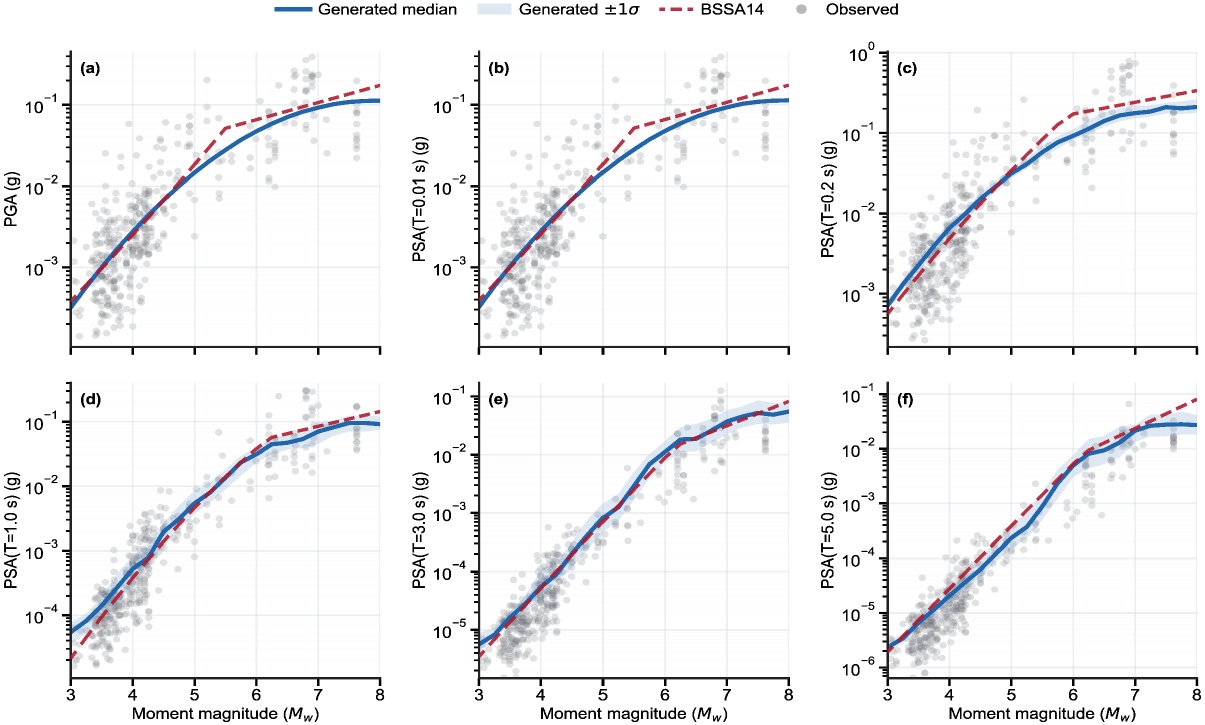}
\caption{Magnitude scaling of RotD50 intensity measures. PGA and multi-period PSA are shown as functions of $M_w$ for strike-slip conditions at $R_{JB}=30$\,km and $V_{S30}=760$\,m/s.}
\label{fig:magnitude_scaling}
\end{figure*}

The magnitude-dependent median curves further test whether the source terms control both amplitude level and spectral-shape migration. At $R_{JB}=30$\,km and $V_{S30}=760$\,m/s under strike-slip conditions, PGA and PSA at 0.01 and 0.2\,s increase rapidly with $M_w$ and show weaker amplitude growth at the large-magnitude end (Figure~\ref{fig:magnitude_scaling}). This weakening of short-period magnitude scaling can be interpreted as magnitude saturation. ASK14 uses segmented magnitude terms and finite-fault simulation constraints for large-magnitude scaling \citep{Abrahamson_2014_SummaryASK14Ground}. The PISL-GMM of Chen et al. also includes magnitude saturation as an explicit physical prior \citep{Chen_2024_PhysicsSymbolicLearner}. In the same scenario, BSSA14 mainly shows reduced magnitude sensitivity at large magnitudes, without necessarily reaching a full plateau \citep{Boore_2014_NGAWest2EquationsPredicting}. By contrast, the PSA ordinates at 1, 3, and 5\,s continue to increase over a wider magnitude range, so the long-period to short-period ratio increases with magnitude. This spectral shift is consistent with the empirical expectation that large events contain a larger fraction of low-frequency energy. It also indicates that the waveform generator preserves period-dependent source scaling beyond the PGA amplitude anchor. The generated medians remain continuous at small magnitudes and fall within the main observed scatter. Although records are sparse at the large-magnitude, long-period end, the flattening of the 5\,s curve remains consistent with the available observations and may indicate long-period magnitude saturation.

\begin{figure*}[!htbp]
\centering
\includegraphics[width=0.8\textwidth]{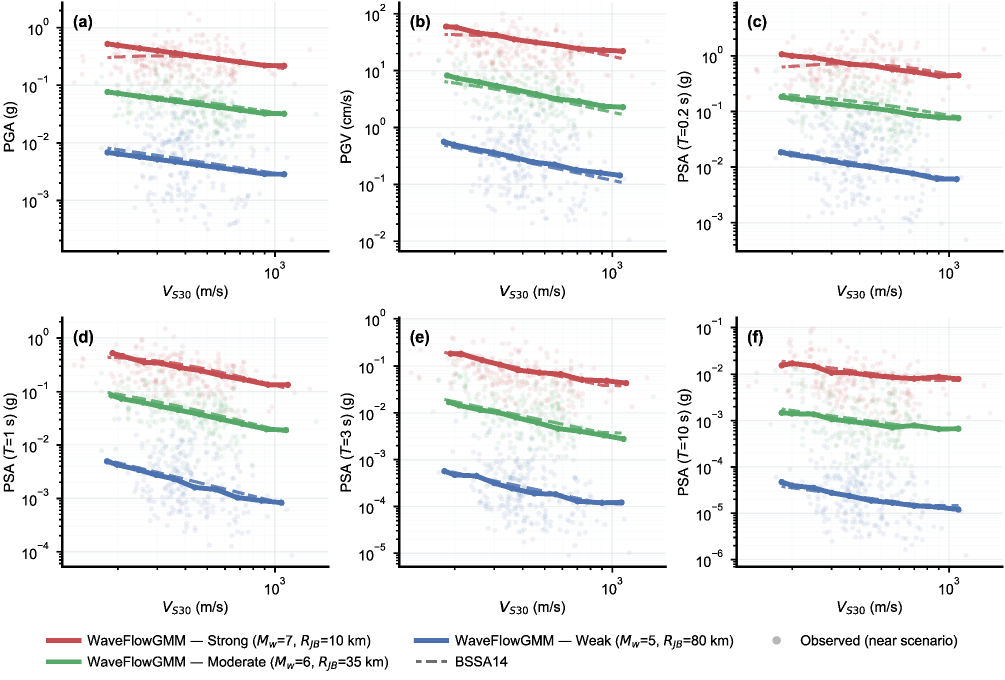}
\caption{Dependence of RotD50 intensity measures on site velocity. PGA, PGV, and PSA are shown as functions of $V_{S30}$ under three magnitude-distance scenarios.}
\label{fig:site_nl}
\end{figure*}

Under strong near-source, moderate, and weak far-field scenarios, PGA, PGV, and PSA at all periods decrease as $V_{S30}$ increases. The amplitude ordering of the three scenarios remains stable over the full $V_{S30}$ range (Figure~\ref{fig:site_nl}). The generated medians for PGA, PGV, and PSA at 0.2--1\,s are close to the BSSA14 reference and fall within the main scatter of nearby observations. This indicates that the PISL amplitude interface and the AlphaFlow waveform generator jointly preserve the short-period site-amplification trend. PSA at 3 and 10\,s is less sensitive to $V_{S30}$, especially towards the high-$V_{S30}$ end. This pattern is consistent with the empirical view that long-period response is less sensitive to shallow $V_{S30}$. Overall, Figure~\ref{fig:site_nl} shows that the generator inherits the $V_{S30}$-dependent amplitude changes from the PISL PGA model and preserves period-dependent site-response changes in the normalised waveform.

\begin{figure*}[!htbp]
\centering
\includegraphics[width=0.8\textwidth]{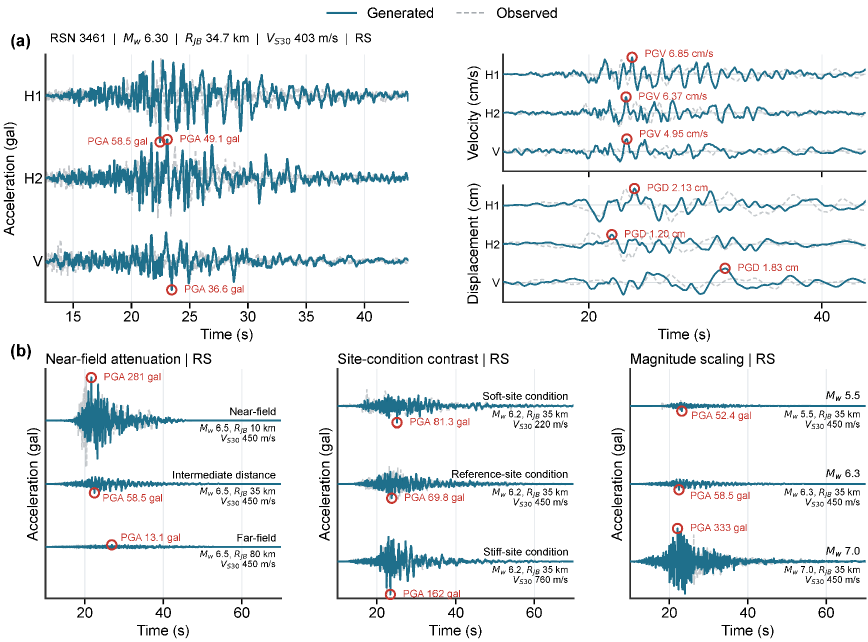}
\caption{Waveform comparisons and generated samples under varying scenario conditions. The upper panels show three-component acceleration, velocity, and displacement, and the lower panels show generated acceleration under distance, site, and magnitude perturbations.}
\label{fig:cond_plate}
\end{figure*}

Figure~\ref{fig:cond_plate} shows that the generated waveforms and a representative test record have similar time-domain behaviour in acceleration, velocity, and displacement. The velocity and displacement time histories of the generated waveform in the figure are obtained by applying a 0.1 Hz, 4th-order zero-phase Butterworth high-pass filter and trapezoidal integration to the acceleration. The three components have comparable strong-motion windows, envelope decay, and velocity and displacement levels, while local phase and peak locations remain stochastic. This is consistent with probabilistic waveform generation: the model learns waveform statistics under observed conditions rather than reconstructing a single record point by point. The condition-perturbation samples show that increasing $R_{JB}$ from 10 to 80\,km reduces peak amplitude and weakens the strongest shaking. Increasing $M_w$ from 5.5 to 7.0 increases both amplitude and effective duration. Under changes in $V_{S30}$, local peaks are more strongly affected by random phase, but the samples retain stable envelopes and plausible frequency content across site conditions. Lower $V_{S30}$ lengthens the near-peak segment and strengthens later oscillations. Thus, Figure~\ref{fig:cond_plate} links the source, path, and site trends in Figures~\ref{fig:distance_scaling}--\ref{fig:site_nl} to the time-domain samples, while preserving record-level randomness.

\subsection{Event residuals and period-dependent dispersion}\label{ssec:residuals}

This section uses between-event and within-event residual diagnostics to check whether systematic bias remains beyond the median-scaling curves. For record $j$ from event $i$, the residuals of any scalar intensity measure $y$ are decomposed as
\begin{equation}
\begin{aligned}
r_{ij} &= \ln y_{\mathrm{obs},ij}-\ln \widehat{y}_{\mathrm{gen},ij},\\
\eta_i &= \frac{1}{n_i}\sum_{j=1}^{n_i}r_{ij},\\
\delta_{ij} &= r_{ij}-\eta_i.
\end{aligned}
\end{equation}
Here, $r_{ij}$ is the natural-log residual of the observation relative to the generated prediction, $\widehat{y}_{\mathrm{gen},ij}$ is the generated prediction under the same condition, $\eta_i$ is the event-mean residual, and $\delta_{ij}$ is the within-event residual after removing the event mean. This treatment follows the between-event and within-event decomposition in the random-effects model of Abrahamson and Youngs \citep{Abrahamson_1992_StableAlgorithmRegression} and is used to diagnose systematic residual trends. The between-event, within-event, and total dispersions are
\begin{equation}
\tau=\sqrt{\mathrm{Var}_i(\eta_i)},\quad
\phi=\sqrt{\mathrm{Var}_{ij}(\delta_{ij})},\quad
\sigma=\sqrt{\tau^2+\phi^2}.
\end{equation}

\begin{figure*}[!htbp]
\centering
\includegraphics[width=0.8\textwidth]{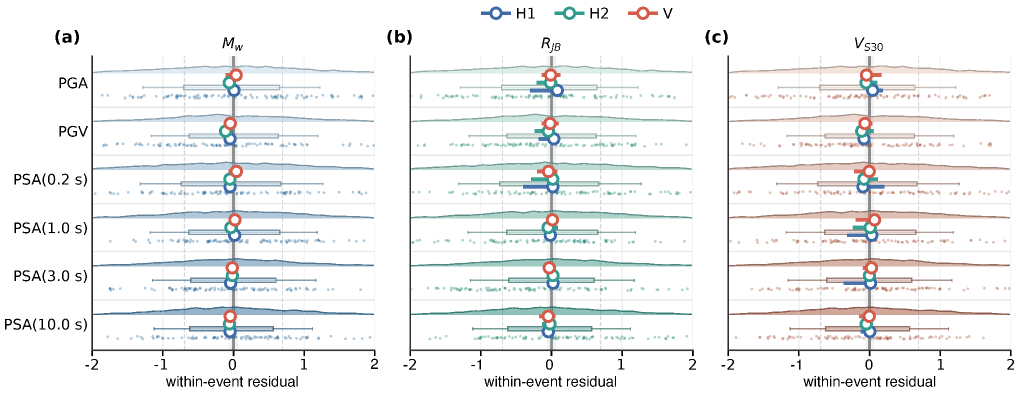}
\caption{Within-event residual distributions of PGA, PGV, and PSA for the three ground-motion components.}
\label{fig:residual_forest_intra}
\end{figure*}

\begin{figure*}[!htbp]
\centering
\includegraphics[width=0.8\textwidth]{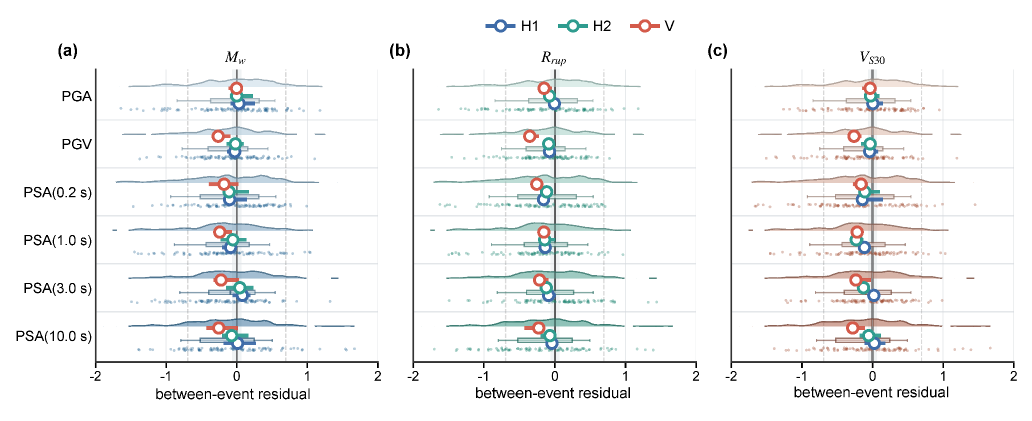}
\caption{Between-event residual distributions of PGA, PGV, and PSA for the three ground-motion components.}
\label{fig:residual_forest_inter}
\end{figure*}

\begin{figure*}[!htbp]
\centering
\includegraphics[width=0.8\textwidth]{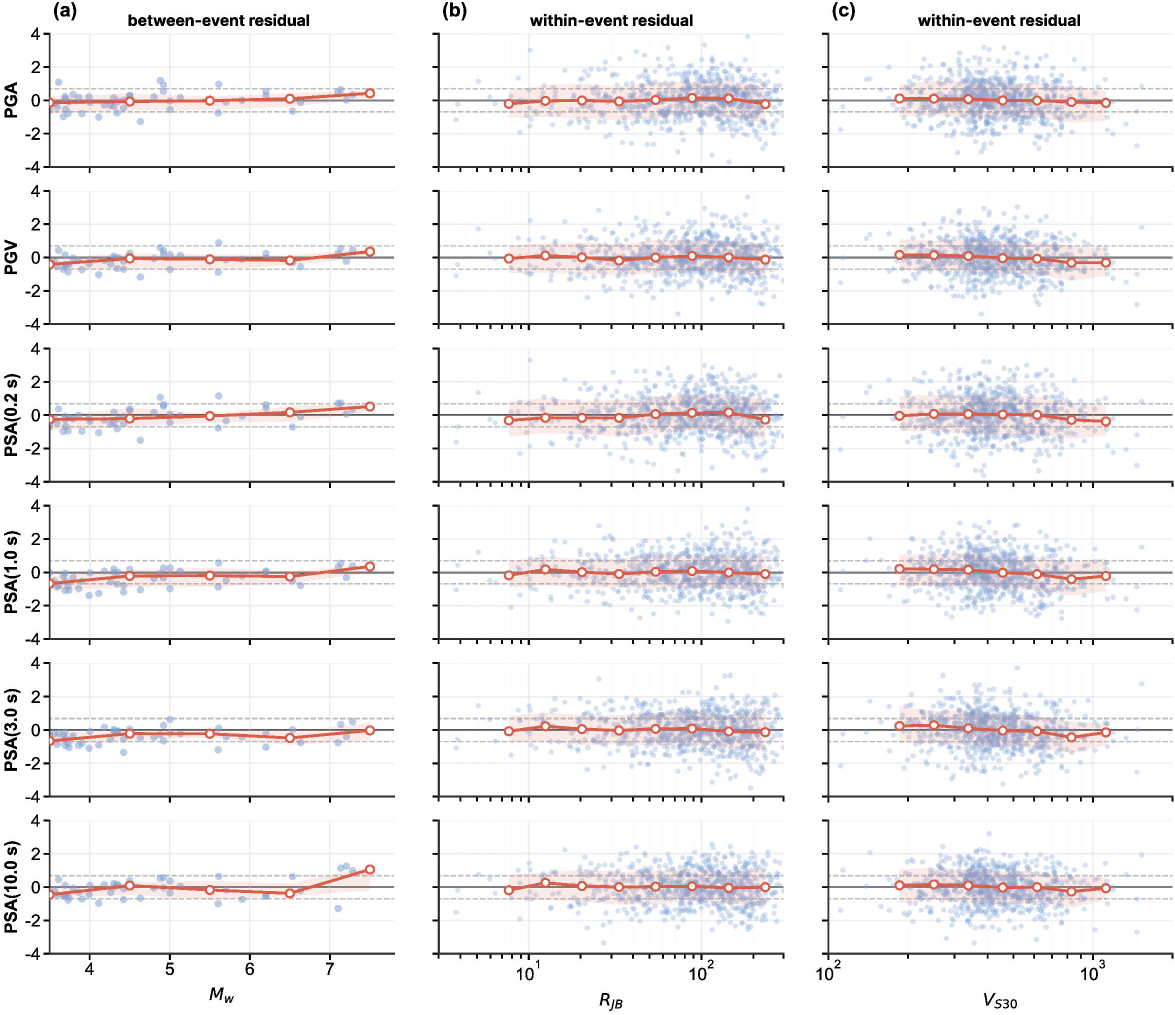}
\caption{H1 residual trends with magnitude, distance, and site velocity.}
\label{fig:resid_combined}
\end{figure*}

The three-component residual-summary plots show that, after binning by $M_w$, $R_{JB}$, and $V_{S30}$, both the event-mean residual $\eta_i$ and within-event residual $\delta_{ij}$ are centred near zero (Figures~\ref{fig:residual_forest_intra} and \ref{fig:residual_forest_inter}). The H1 three-axis diagnostic further shows no persistent monotonic trend with magnitude, distance, or site condition. Most binned medians and their 16th-to-84th-percentile envelopes fall within the $\pm\ln2$ reference band marked by the grey dashed lines (Figure~\ref{fig:resid_combined}). Visible offsets mainly occur beyond 200\,km for PSA at 3 and 10\,s, matching the long-period, far-distance region with the weakest data coverage.
The corresponding H2 and vertical-component residual diagnostics are provided in the Supplementary Materials.

\begin{figure*}[!htbp]
\centering
\includegraphics[width=0.8\textwidth]{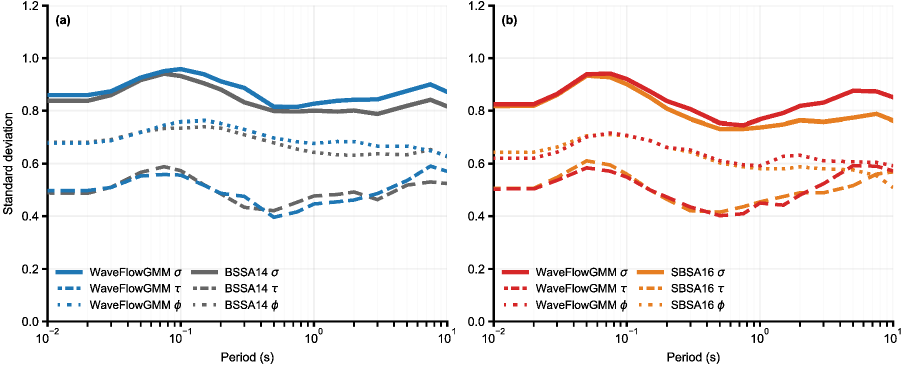}
\caption{Period dependence of PSA residual dispersion for WaveFlowGMM and reference ground-motion models.}
\label{fig:residual_vs_period}
\end{figure*}

Figure~\ref{fig:residual_vs_period} shows the period-dependent residual variability of the generated waveforms, with WaveFlowGMM evaluated on the event-level holdout set. For each test record, 50 waveforms are generated under identical conditions, and the sample-median PSA is used as the model prediction. The residual is defined as $\ln\mathrm{PSA}_{\mathrm{obs}} - \ln\widetilde{\mathrm{PSA}}_{\mathrm{gen}}$. Thus, the plotted $\sigma$, $\tau$, and $\phi$ represent the between-event and within-event dispersion of the sample-median prediction error, not the full aleatory variability of a single random generated sample. BSSA14 and SBSA16 \citep{Stewart_2016_NGAWest2EquationsPredicting} are fixed empirical equations, and their baseline residuals are computed with the actual scenario parameters of each record. All records contribute only at spectral periods supported by their lowest usable frequency, excluding long-period filtering noise from the dispersion estimate.

As shown in Figure~\ref{fig:residual_vs_period}, the total $\sigma$ curve of the generated horizontal component is close to BSSA14 and is slightly higher near the short-period peak and at the long-period end. For the vertical component, WaveFlowGMM is also close to SBSA16, with the main difference concentrated around the short-period peak. In both components, the dispersion is dominated by the within-event term $\phi$, while the period dependence of $\tau$ follows a shape similar to the empirical baselines. This result indicates that the sample-median waveform prediction does not produce anomalously low or high residual dispersion, and that its period dependence is broadly consistent with empirical GMMs.

\subsection{Few-step sampling and backbone comparison}\label{ssec:efficiency}

\begin{figure*}[!htbp]
\centering
\includegraphics[width=0.8\textwidth]{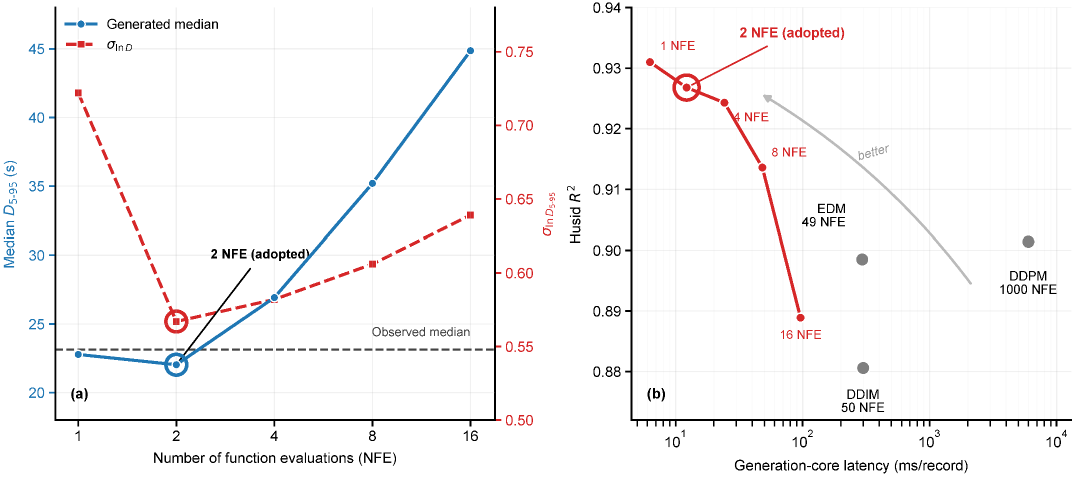}
\caption{Sampling-step sensitivity and generation efficiency. (a) Median duration and logarithmic duration residual standard deviation versus NFE. (b) Generation time versus Husid-curve $R^2$ for AlphaFlow and diffusion baselines.}
\label{fig:efficiency}
\end{figure*}

To evaluate the performance of the AlphaFlow base few-step configuration in ground motion generation, Table~\ref{tab:nfe} and Figure~\ref{fig:efficiency}a compare the sampling results of the same fully trained AlphaFlow model under 1 to 16 NFE. The current timing setup does not include a separate GPU warm-up stage, so the first batch of inference may include one-time initialisation overhead. The generation core time is defined as the sum of three components: PGA sampling, AlphaFlow network sampling, and inverse wavelet packet reconstruction. The per-record time is obtained by dividing the batch generation core time by the number of records in that batch. Dataset construction, data loading, model weight loading, input preprocessing, forward wavelet packet transform, and disk writing are not included in the generation core time.

The 2-NFE configuration gives the most favourable combination of duration diagnostics and computational cost within the settings examined here. The generated $D_{5\text{-}95}$ median is 22.03\,s, close to the observed median of 23.13\,s; the record-level log-duration correlation $r=0.709$ is the highest; and the duration residual standard deviation $\sigma=0.567$ is the lowest. With 1 NFE, the median remains close to the observation, but the correlation falls to 0.492 and the dispersion rises to 0.722, indicating that one step under-resolves the present velocity field. With 4 to 16 NFE, the generated median increases from 26.92 to 44.86\,s, while correlation and dispersion both degrade, producing overly long duration. This indicates that increasing the number of sampling steps does not continuously improve the duration accuracy of the current model.

This non-monotonicity arises from the AlphaFlow training objective: training anneals from trajectory flow matching towards an average-velocity field, so the network learns the average velocity over a given step rather than the instantaneous velocity. Too few steps under-resolve the velocity field, whereas too many small steps drift away from the average-velocity trajectory emphasised during training and accumulate error, systematically lengthening duration. We selected 2-NFE based on duration statistics and computational cost on the evaluation set of 54 events. Because this set was used for both configuration selection and performance evaluation, the reported performance may be optimistic.

\begin{table*}[!htbp]
\centering
\caption{Duration accuracy and generation time in the AlphaFlow NFE sensitivity analysis.}
\label{tab:nfe}
\begin{tabular}{rccccc}
\toprule
NFE & Gen. median (s) & Obs. median (s) & $\ln D_{5\text{-}95}$ $r$ & $\ln D_{5\text{-}95}$ $\sigma$ & Generation time (ms/record) \\
\midrule
1 & 22.78 & 23.13 & 0.492 & 0.722 & 6.1 \\
\textbf{2} & \textbf{22.03} & \textbf{23.13} & \textbf{0.709} & \textbf{0.567} & \textbf{12.2} \\
4 & 26.92 & 23.13 & 0.691 & 0.582 & 24.4 \\
8 & 35.20 & 23.13 & 0.653 & 0.606 & 48.8 \\
16 & 44.86 & 23.13 & 0.613 & 0.639 & 96.1 \\
\bottomrule
\end{tabular}
\end{table*}

To demonstrate the advantage of the AlphaFlow framework used in this study, we conduct an ablation across different generative models. The wavelet-packet target, four-condition interface, PGA-derived flow condition, training data, and training duration are fixed, while only the generative backbone and sampler change. The comparison includes DDPM \citep{Ho_2020_DenoisingDiffusionProbabilistic}, DDIM \citep{Song_2022_DenoisingDiffusionImplicit}, EDM \citep{Karras_2022_ElucidatingDesignSpace}, and AlphaFlow. The four backbones have similar amplitude accuracy (Table~\ref{tab:backbone}, Figure~\ref{fig:efficiency}b): H1 PGA correlations are 0.880 to 0.887, and PGA dispersions are 1.25 to 1.28 natural-log units. AlphaFlow gives the highest PSA(1\,s) correlation and a Husid $R^2$ of 0.927, compared with 0.881 to 0.901 for the diffusion samplers. Its generation-core runtime is 12.2\,ms per record, about 1/24 of the 50-step DDIM and 49-evaluation EDM samplers and about 1/490 of the 1000-step DDPM ancestral sampler.

Under the hardware and batch settings in this paper, AlphaFlow significantly reduces the core time cost for generating a single three-component waveform, lowering the generation cost of a single waveform to the order of one percent of that of the diffusion sampler. This result indicates that the model is suitable for repeated conditional waveform sampling for multiple scenarios and large numbers of samples, and provides support in terms of computational efficiency for embedding conditional waveform generation into waveform-level PSHA workflows.

\begin{table*}[!htbp]
\centering
\caption{Waveform metrics and generation time of generative backbones.}
\label{tab:backbone}
\begin{tabular}{lrccccc}
\toprule
Backbone & NFE & PGA $r$ & PGA $\sigma$ & PSA(1\,s) $r$ & Husid $R^2$ & Generation time (ms/record) \\
\midrule
DDPM \citep{Ho_2020_DenoisingDiffusionProbabilistic}  & 1000 & 0.880 & 1.279 & 0.893 & 0.901 & 5990.3 \\
DDIM \citep{Song_2022_DenoisingDiffusionImplicit} & 50   & 0.882 & 1.267 & 0.894 & 0.881 & 299.6 \\
EDM \citep{Karras_2022_ElucidatingDesignSpace}        & 49   & 0.887 & 1.245 & 0.885 & 0.899 & 294.4 \\
\textbf{AlphaFlow (this work)}                        & \textbf{2} & \textbf{0.883} & \textbf{1.266} & \textbf{0.899} & \textbf{0.927} & \textbf{12.2} \\
\bottomrule
\end{tabular}
\end{table*}

\section{Discussion}\label{sec:discussion}

Empirical GMMs have developed a relatively mature low-dimensional amplitude prediction framework, whereas full waveform synthesis remains a complex high-dimensional conditional distribution modelling problem. WaveFlowGMM treats these two problems separately, enabling the errors of the amplitude model and the waveform generator to be diagnosed independently. PGA normalisation separates the absolute amplitude scale from the generation target, but the three-component PGA is still fed to the generator as an explicit conditional input, enabling it to learn how waveform morphology and relative energy distribution vary with ground motion intensity. Nevertheless, PGA cannot uniquely determine the spectrum, duration, and cumulative energy, and the systematic biases in PGV, PGD, Arias intensity, and CAV also indicate that there is still room for improvement in the normalised waveform distribution. Joint conditioning on multiple IMs and its end-to-end validation will be pursued in future research.

Multicomponent engineering analyses depend on the relative amplitudes of the three-component waveforms, and these ratios are shaped by source radiation, path geometry, and site response acting on the same record. A model that samples the three components as independent marginal amplitudes loses this dependence even if each marginal distribution is reasonable. The full covariance therefore turns the PGA stage from three separate scalar predictors into a joint random variable. This distinction matters for multicomponent record selection, vertical-to-horizontal response studies, and waveform-level hazard sampling where component ratios are carried into structural response.

Few-step AlphaFlow sampling makes waveform-level use computationally plausible, but the appropriate step count remains an engineering choice. The results show that peak-amplitude agreement alone is not enough to select a sampler. Duration, Arias energy build-up, velocity, displacement, and long-period response are the quantities that reveal whether a generated time history remains useful after integration and structural filtering. Under the specified filtering and integration procedure, the generated three-component acceleration can also be integrated into velocity and displacement without showing an obvious systematic trend, providing an additional check on the low-frequency content of the samples. For probabilistic seismic hazard analysis, this framework is best understood as a candidate waveform-level component. A complete deployment still requires a standardised hazard integral and validation against engineering demand parameters.

WaveFlowGMM describes inter-record variability under a given scenario through PGA residual sampling and waveform-generated noise. The scatter of generated samples captures conditional variation under the fixed model, including source, path, and site differences unresolved by the current conditioning variables. Limited NGA-West2 coverage, model form, and parameter estimation introduce epistemic uncertainty, especially in sparsely sampled scenarios. This uncertainty is not quantified separately here, so the generated quantile intervals describe conditional spread under the fixed model. A complete uncertainty assessment requires further analysis.

In addition, the PISL site term is a linear saturated $V_{S30}$ ratio and does not model nonlinear soil response that couples $V_{S30}$ with input intensity. The model generates aligned waveform windows, not absolute arrival times, so network-level timing applications remain outside the present scope. NGA-West2 remains sparse in the large-magnitude, near-source, and long-period regimes where both empirical GMMs and data-driven generators are hardest to constrain. The NGA-West3 programme is expanding and updating crustal ground-motion databases for the next generation of GMMs. Such larger and more diverse datasets should allow future versions of waveform-level GMMs to learn sharper conditional distributions, reduce smoothing in data-poor regions, and test regional transferability more directly.

\section{Conclusions}\label{sec:conclusion}

This study introduced WaveFlowGMM, a two-stage ground-motion model for scenario-conditioned generation of three-component acceleration histories. Its main design is a probabilistic PGA interface with full cross-component covariance, coupled to a few-step AlphaFlow generator in reconstructible wavelet-packet space. Tests on the event-level NGA-West2 holdout set showed that the generated histories recover the main magnitude, distance, and site scaling, with peak and spectral residuals close to zero and dispersion comparable to empirical GMMs. The full-covariance PGA model reproduced the measured three-component PGA residual correlations, with coefficients of 0.934 for H1\textendash H2, 0.860 for H1\textendash V, and 0.860 for H2\textendash V. Under the hardware and comparison settings used in this study, the 2-NFE AlphaFlow required 12.2 ms per record for core generation, reducing the per-record cost to the order of one percent of that of the diffusion samplers.

These results support WaveFlowGMM as a candidate component for waveform-level seismic hazard and risk workflows. Its current scope remains bounded by NGA-West2 coverage and the conditioning variables used here. Larger regional databases and richer rupture-path-site descriptors are the next step towards site-specific waveform-level analysis.

\section*{Acknowledgements}

This work was supported by the National Natural Science Foundation of China (No. 42574127), the Special Fund of the Institute of Geophysics, China Earthquake Administration (No. DQJB26A49), and the 2026 Consulting Research Project of the Wuhan Branch of the Hubei Research Institute for the Development Strategy of China Engineering Science and Technology (No. HB2026B14). The authors thank the Pacific Earthquake Engineering Research Center for providing the NGA-West2 database used in this study. This project is supported by the Intelligent Computing Center of Jianghan University.

\section*{Conflict of Interest Statement}

The authors declare no conflicts of interest.

\section*{Data Availability Statement}

The NGA-West2 database used in this study is available from the Pacific Earthquake Engineering Research Center (https://ngawest2.berkeley.edu/, last accessed May 2024).

The code will be publicly available on GitHub at \url{https://github.com/CyrusMagic/WaveFlowGMM} upon publication.

\appendix

\section{Notation}\label{app:notation}

The main symbols used throughout Section~2 are summarised in Table~\ref{tab:notation}.

\begin{table*}[!htbp]
\centering

\caption{Notation used in the conditional amplitude and waveform-generation framework.}
\label{tab:notation}
\renewcommand{\arraystretch}{1.12}
\begin{tabularx}{\textwidth}{@{}l l X@{}}
\toprule
Symbol & Dimension or representation & Definition \\
\midrule
$\mathbf{x}$ & -- & Scenario condition vector containing $M_w$, $R_{JB}$, $V_{S30}$, and faulting mechanism. \\
$\mathbf{q}$ & $3$ & Base-10 logarithms of the H1, H2, and V PGA values. \\
$\mathbf{r}(\mathbf{q})$ & $3$ & Compact PGA condition containing the horizontal amplitude level and the H1/H2 and V-to-horizontal amplitude ratios. \\
$\boldsymbol{\Sigma}_{\tau}$, $\boldsymbol{\Sigma}_{\phi}$ & $3\times3$ & Between-event and within-event covariance matrices of the three-component PGA residuals. \\
$p_{\psi}$, $p_{\theta}$ & Probability distributions & Conditional distribution of the three-component PGA; conditional distribution of the normalised waveform. \\
$\boldsymbol{\mu}(\mathbf{x})$, $\beta_{d,k}$ & -- & Conditional mean of the component-wise log-PGA; fitted coefficients of the PISL amplitude model. \\
$\mathbf{w}_0$, $\mathbf{w}_t$, $\mathbf{w}$ & Wavelet-packet coefficient space & $\mathbf{w}_0$ is the coefficient representation of the PGA-normalised waveform; $\mathbf{w}_t$ is the intermediate state at flow time $t$; $\mathbf{w}$ is the current state in Algorithm~1. \\
$\mathbf{a}$, $\tilde{\mathbf{a}}$, $\hat{\mathbf{a}}$ & Acceleration time history & $\mathbf{a}$ is the observed acceleration; $\tilde{\mathbf{a}}$ is the inverse-transformed normalised acceleration; $\hat{\mathbf{a}}$ is the final generated acceleration after PGA restoration. \\
$\mathbf{v}_t$, $\mathbf{u}_{\theta}$, $\mathbf{v}^{\mathrm{tgt}}_t$ & Coefficient-space velocity & $\mathbf{v}_t$ is the instantaneous path velocity; $\mathbf{u}_{\theta}$ is the predicted conditional average velocity; $\mathbf{v}^{\mathrm{tgt}}_t$ is the training target velocity. \\
$t$, $\Delta t$ & -- & Normalised flow time and flow integration step. \\
$K$ or NFE & -- & Number of AlphaFlow integration steps or neural function evaluations. \\
$\alpha$, $s$ & -- & AlphaFlow training-objective mixing parameter and training progress. \\
\bottomrule
\end{tabularx}
\end{table*}

% BEGIN reviewer 1 Appendix B revision

\clearpage

\begingroup

\section{Distribution-based validation and scenario-stratified analysis of generated ground motions}\label{app:distribution}

\setcounter{table}{0}

\renewcommand{\thetable}{B\arabic{table}}

\renewcommand{\theHtable}{B.\arabic{table}}

\setcounter{figure}{0}

\renewcommand{\thefigure}{B\arabic{figure}}

\renewcommand{\theHfigure}{B.\arabic{figure}}

% MD retains variance-ratio prose although its Table B1 omits V; reconcile before submission.

% The specified Figure B1 PDF is used as supplied; reconcile its values with the final source data.

\subsection{Multi-sample generation and probabilistic evaluation}\label{app:probabilistic}

This appendix supplements Figure~\ref{fig:im_density} with distributional statistics and interval coverage. The test set, partitioned by earthquake event, contains 3,143 records from 54 events. Magnitude, distance, site velocity, and faulting mechanism are held fixed for each record. Resampling three-component PGA and waveform-generation noise produces 50 three-component waveforms per scenario. Sampling uses two NFE, and the IM analysis uses H1. The same IM definitions, physical units, and record windows are used for observed and generated waveforms.

Table~\ref{tab:appendix_probabilistic} compares distributions across all test scenarios and reports predictive scores calculated within each scenario. For each IM, quantiles and logarithmic variances are computed separately for the observed and generated samples. Their generated-to-observed ratios and the standardised Wasserstein distance describe the distributions. CRPS and interval coverage are calculated within each scenario and then averaged over records. The ratios $R_{0.05}$, $R_{0.50}$, and $R_{0.95}$ compare corresponding generated and observed quantiles. Quantiles are obtained by linear interpolation in the $\log_{10}$ domain and then transformed back to the original scale. The ratio $V$ compares the variances of the logarithmic IM values, with each variance normalised by its sample count. The Wasserstein distance $W_1^*$ is calculated between the empirical distributions of logarithmic IM values and divided by the observed logarithmic standard deviation.

For each scenario, the generated IM samples are evaluated against the corresponding observation. The continuous ranked probability score (CRPS) combines the distances from generated values to the observation with the dispersion among generated samples. Let $x_i$ be the logarithmic observation, $x_{ij}$ a logarithmic generated value, and $s_{\mathrm{obs}}$ the observed logarithmic standard deviation. The mean empirical CRPS is standardised as follows:

\begin{equation*}
\begin{aligned}
\mathrm{CRPS}^* &= \frac{1}{Ns_{\mathrm{obs}}}\sum_{i=1}^{N}
\biggl[\frac{1}{K}\sum_{j=1}^{K}|x_{ij}-x_i|\\
&\qquad -\frac{1}{2K^2}\sum_{j=1}^{K}\sum_{l=1}^{K}|x_{ij}-x_{il}|\biggr].
\end{aligned}
\end{equation*}

Here, $N$ is the number of test records and $K$ is the number of generated samples per scenario.

Coverage measures the proportion of observations falling within the generated intervals for their corresponding scenarios. For each scenario, the 5th and 95th percentiles of generated logarithmic IM values define the interval boundaries. Each observation is checked against its corresponding interval, and the results are aggregated as a proportion of records. This interval reflects uncertainty due to sampling test events with the generated samples held fixed.

\begin{table*}[!t]
\centering

\caption{Distributional and probabilistic evaluation of H1 intensity measures.}
\label{tab:appendix_probabilistic}
\small
\setlength{\tabcolsep}{6pt}
\renewcommand{\arraystretch}{1.1}
\begin{tabular}{@{}lrrrrrl@{}}
\toprule
IM & $R_{0.05}$ & $R_{0.50}$ & $R_{0.95}$ & $W_1^*$ & $\mathrm{CRPS}^*$ & $C_{90}$ [95\% CI] (\%) \\
\midrule
PGA & 1.05 & 0.89 & 0.91 & 0.086 & 0.210 & 87.3 [83.9, 90.8] \\
PGV & 1.12 & 1.13 & 1.16 & 0.077 & 0.175 & 88.1 [85.7, 90.7] \\
PGD & 1.22 & 1.54 & 1.21 & 0.128 & 0.172 & 80.3 [76.0, 84.9] \\
Arias intensity & 1.29 & 1.28 & 1.18 & 0.086 & 0.187 & 88.9 [86.0, 92.0] \\
CAV & 1.31 & 1.37 & 1.21 & 0.122 & 0.188 & 86.9 [82.9, 90.6] \\
EPA & 1.07 & 0.95 & 0.98 & 0.096 & 0.209 & 88.4 [85.4, 91.5] \\
$\mathrm{PSA}_{\max}$ & 1.14 & 0.79 & 0.85 & 0.097 & 0.224 & 86.6 [82.8, 90.3] \\
$\mathrm{PSV}_{\max}$ & 1.21 & 1.22 & 0.92 & 0.069 & 0.177 & 86.8 [84.4, 89.2] \\
$\mathrm{PSD}_{\max}$ & 1.33 & 1.21 & 0.78 & 0.111 & 0.153 & 80.4 [75.9, 83.6] \\
SI & 1.23 & 1.27 & 1.15 & 0.127 & 0.173 & 83.1 [78.4, 87.9] \\
$I_c$ & 1.38 & 1.22 & 1.20 & 0.093 & 0.181 & 88.5 [85.2, 91.7] \\
$D_{5\text{-}95}$ & 1.47 & 1.15 & 0.87 & 0.249 & 0.405 & 66.4 [62.5, 69.9] \\
\bottomrule
\end{tabular}
\end{table*}

Quantile agreement and interval coverage vary among IMs (Table~\ref{tab:appendix_probabilistic}). Coverage is 88.1\% for PGV and 88.9\% for Arias intensity, both close to the nominal 90\% level. Their median ratios of 1.13 and 1.28 show that this coverage coexists with a shift in the distributions. Coverage falls to 80.3\% for PGD and 80.4\% for $\mathrm{PSD}_{\max}$, and both 95\% confidence intervals lie below 90\%. 

Duration has the largest distributional bias. Its 5th- and 95th-percentile ratios are 1.47 and 0.87, respectively, with both tails shifted towards the centre. It also has the highest standardised Wasserstein distance and CRPS in Table~\ref{tab:appendix_probabilistic}, and coverage is only 66.4\%. The discrepancies affect both the distribution and predictions for individual scenarios. Appendix~\ref{app:stratified} examines how they vary across scenario ranges.

\subsection{Stratified evaluation of intensity-measure distributions}\label{app:stratified}

The test set is stratified separately by magnitude, distance, and site velocity, using thresholds of $M_w=4$ and 6, $R_{JB}=50$ and 120 km, and $V_{S30}=300$ and 500 m/s. Figure~\ref{fig:appendix_distribution} compares PGV, Arias intensity, and significant duration through their 5th-percentile ratios, 95th-percentile ratios, and 90\% interval coverage. Coverage is the proportion of records whose observations fall within their corresponding scenario intervals, as defined in Appendix~\ref{app:probabilistic}.

Duration shows narrower distributions and insufficient coverage in every stratum. Its 5th-percentile ratios range from 1.16 to 2.13, its 95th-percentile ratios from 0.82 to 0.97, and its coverage from 60.4\% to 68.4\%. These differences occur across magnitude, distance, and site-velocity ranges.

Coverage ranges from 84.1\% to 92.4\% for PGV and from 85.1\% to 93.5\% for Arias intensity. Both are higher overall than duration coverage but vary across strata. For $V_{S30}\leq300$ m/s, coverage is 92.4\% and 93.5\%, respectively. For $V_{S30}>500$ m/s, it is 84.2\% and 85.1\%, respectively. Arias-intensity 95th-percentile ratios range from 0.82 to 1.34, also showing pronounced differences in upper-tail bias across strata.

Table~\ref{tab:appendix_pga_psa_stratified} extends this analysis to PGA and 5\%-damped PSA at 1, 3, and 10\,s using the same strata. Records with missing or invalid usable frequencies are excluded from PSA comparisons. The same record selection is applied to observations and their corresponding generated samples.

\begin{figure*}[!t]
\centering
\includegraphics[width=0.9\textwidth]{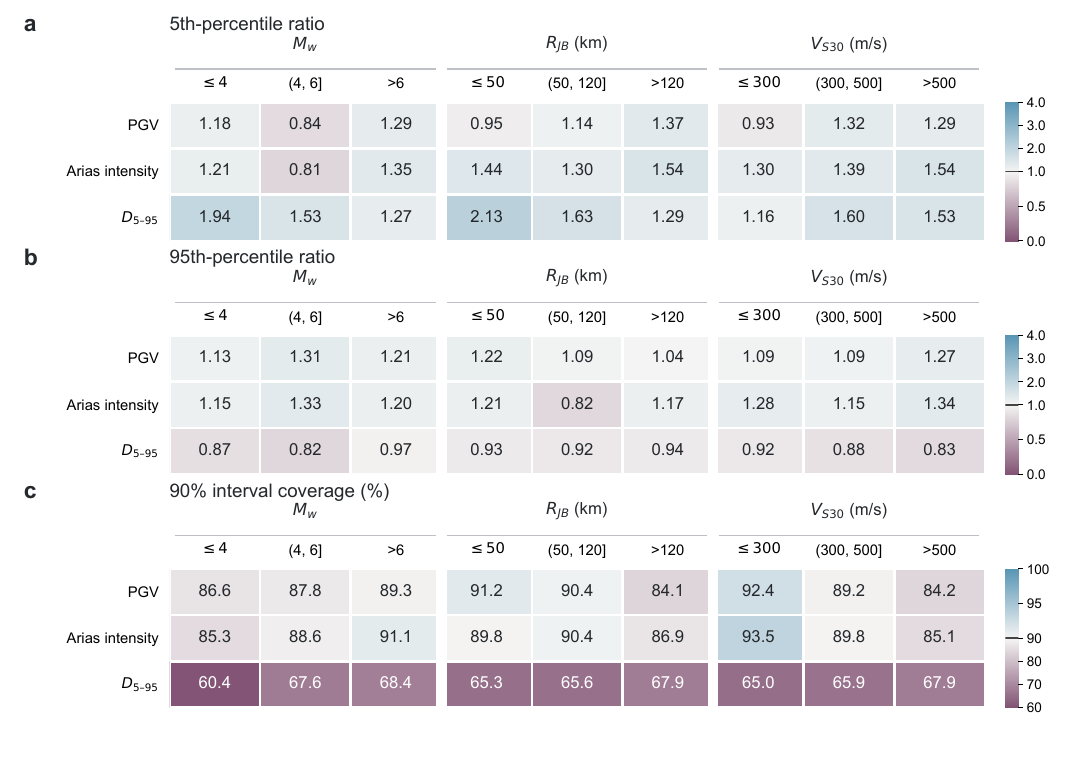}
\caption{Distributional comparisons and interval coverage of H1 intensity measures across magnitude, distance, and site-velocity strata.}
\label{fig:appendix_distribution}
\end{figure*}

\begin{table*}[!t]
\centering

\caption{Stratified distributional comparisons and interval coverage of H1 PGA and representative PSA ordinates.}
\label{tab:appendix_pga_psa_stratified}
\small
\setlength{\tabcolsep}{3pt}
\renewcommand{\arraystretch}{1.15}
\begin{tabular*}{\textwidth}{@{\extracolsep{\fill}}l*{12}{r}@{}}
\toprule
\multirow{2}{*}{Stratum} & \multicolumn{3}{c}{PGA} & \multicolumn{3}{c}{PSA(1\,s)} & \multicolumn{3}{c}{PSA(3\,s)} & \multicolumn{3}{c}{PSA(10\,s)} \\
\cmidrule(lr){2-4}\cmidrule(lr){5-7}\cmidrule(lr){8-10}\cmidrule(lr){11-13}
 & $R_{0.05}$ & $R_{0.95}$ & $C_{90}$ (\%) & $R_{0.05}$ & $R_{0.95}$ & $C_{90}$ (\%) & $R_{0.05}$ & $R_{0.95}$ & $C_{90}$ (\%) & $R_{0.05}$ & $R_{0.95}$ & $C_{90}$ (\%) \\
\midrule
\multicolumn{13}{@{}l}{Magnitude} \\
$M_w\leq4$ & 1.00 & 1.00 & 86.2 & 1.24 & 1.31 & 86.9 & 1.32 & 1.38 & 83.4 & 1.47 & 0.88 & 80.8 \\
$4<M_w\leq6$ & 1.13 & 1.09 & 85.5 & 1.22 & 1.36 & 83.8 & 1.16 & 1.75 & 89.2 & 0.98 & 1.23 & 91.4 \\
$M_w>6$ & 1.11 & 1.06 & 89.7 & 1.37 & 1.20 & 83.9 & 1.03 & 1.04 & 87.3 & 1.20 & 0.84 & 82.5 \\
\addlinespace[3pt]
\multicolumn{13}{@{}l}{Distance (km)} \\
$R_{JB}\leq50$ & 1.14 & 1.16 & 89.0 & 1.28 & 1.21 & 86.4 & 1.25 & 1.30 & 88.9 & 1.06 & 0.87 & 88.6 \\
$50<R_{JB}\leq120$ & 1.07 & 0.82 & 87.0 & 1.12 & 1.10 & 85.0 & 1.25 & 1.20 & 88.7 & 1.14 & 0.77 & 87.7 \\
$R_{JB}>120$ & 1.09 & 0.98 & 86.7 & 1.36 & 1.04 & 83.7 & 1.26 & 1.15 & 87.9 & 1.24 & 0.70 & 81.7 \\
\addlinespace[3pt]
\multicolumn{13}{@{}l}{Site velocity (m/s)} \\
$V_{S30}\leq300$ & 1.04 & 1.09 & 89.9 & 1.12 & 1.27 & 88.2 & 0.83 & 1.05 & 88.5 & 0.93 & 0.79 & 82.1 \\
$300<V_{S30}\leq500$ & 1.11 & 0.86 & 88.3 & 1.21 & 1.19 & 82.8 & 1.19 & 1.15 & 83.8 & 1.18 & 0.80 & 80.5 \\
$V_{S30}>500$ & 1.03 & 0.97 & 84.5 & 1.56 & 1.50 & 78.2 & 1.42 & 1.38 & 81.8 & 1.48 & 0.78 & 84.3 \\
\bottomrule
\end{tabular*}
\par\smallskip
\parbox{\textwidth}{\footnotesize\textit{Note:} $R_{0.05}$ and $R_{0.95}$ are generated-to-observed quantile ratios; $C_{90}$ is the proportion of observations within their corresponding generated 5th--95th-percentile intervals.}
\end{table*}

\clearpage

\endgroup

% END reviewer 1 Appendix B revision

\bibliographystyle{elsarticle-num}
\bibliography{references}

\end{document}